# Why did Einstein Reject the November Tensor in 1912-1913, only to Come Back to it in November 1915?

Galina Weinstein

Department of Philosophy, University of Haifa

### Abstract

The question of Einstein's rejection of the November tensor is re-examined in light of conflicting answers by several historians. I discuss these conflicting conjectures in view of three questions that should inform our thinking: Why did Einstein reject the November tensor in 1912, only to come back to it in 1915? Why was it hard for Einstein to recognize that the November tensor is a natural generalization of Newton's law of gravitation? Why did it take him three years to realize that the November tensor is not incompatible with Newton's law? I first briefly describe Einstein's work in the Zurich Notebook. I then discuss a number of interpretive conjectures formulated by historians and what may be inferred from them. Finally, I offer a new combined conjecture that answers the above questions.

## Contents

## 1. Introduction

Einstein detailed his struggles with mathematical tools that his loyal friend from school, Marcel Grossmann, brought him, in a small notebook from the winter of 1912–1913 – named by scholars the "Zurich Notebook". Grossmann's name was written on top of one of the pages, where Einstein considered candidate field equations with a gravitational tensor constructed from the Ricci tensor; an equation Einstein would return to in his first November 1915 paper on general relativity. However, as indicated in the *Zurich Notebook*, he finally chose non-covariant field equations.

Without going into too many mathematical details here, I would like to present the basic problem. On page 19L Einstein constructed field equations out of the Ricci tensor in first-order approximation. These equations were written again on page 19R:

$$(1)\ \Box g_{ik} = \kappa \rho_0 \frac{dx_i}{d\tau}\frac{dx_k}{d\tau} = \kappa T_{ik}.$$

The left-hand side of these weak-field equations included a reduced Ricci tensor, reduced to d'Alembertian operator in the weak-field limit:

$$\Box = \left(\frac{\partial^2}{\partial x^2} + \frac{\partial^2}{\partial y^2} + \frac{\partial^2}{\partial z^2} - \frac{1}{c^2}\frac{\partial^2}{\partial t^2}\right),$$

while the right-hand side included the contravariant stress-energy tensor $T_{ik}$ for a cloud of pressureless dust multiplied by the gravitational constant $\kappa$.

On page 19L, Einstein introduced the linearized harmonic coordinate condition which was related with compatibility of the field equations and the Newtonian limit (Einstein 1912a, 433-437):

$$(2)\ \sum_\kappa \gamma_{\kappa\kappa}\left(2\frac{\partial g_{i\kappa}}{\partial x_\kappa} - \frac{\partial g_{\kappa\kappa}}{\partial x_i}\right) = 0.$$

The harmonic condition (2) was used to reduce the Ricci tensor to the d'Alembertian acting on the metric tensor field (metric), metric tensor, in the weak-field case, equation (1).

On page 19R, Einstein checked energy and momentum conservation for the gravitational field equations (1) in the case of weak-fields. However, he then realized that he needed an additional condition. He began by writing that for the first-order approximation our additional condition was obtained from the above harmonic coordinate condition. He suggested that the harmonic coordinate condition could perhaps be decomposed into two extra conditions. The first condition is the following:

$$(2a)\ \sum \gamma_{\kappa\kappa}\frac{\partial g_{i\kappa}}{\partial x_\kappa} = 0.$$

This condition is called the "Hertz condition" because it was later mentioned by Einstein in a letter to Paul Hertz (Einstein to Paul Hertz, August 22, 1915, *CPAE* 8, Doc. 111; Renn and Sauer, 2007, 184; Norton 1984, 275).

Einstein imposed the (linearized) Hertz condition to ensure that his field equations (1) were compatible with energy-momentum conservation in first-order approximation. Consequently, the Hertz condition (2a) was added to make sure that the divergence of the stress-energy tensor vanishes in the weak-field case.

Einstein then realized that the combination of the harmonic coordinate condition and the Hertz condition caused a great problem because the trace of the weak-field metric is constant. Thus, the second condition, $\sum_\kappa \gamma_{\kappa\kappa} g_{\kappa\kappa} = const$, was a condition on the trace of the weak-field metric and it was incompatible with Einstein's conception of weak static gravitational fields: a spatial flat metric. Einstein's earlier work on static gravitational fields led him to conclude in the *Zurich Notebook* that in the weak-field approximation, the spatial metric of a static gravitational field must be flat (Einstein 1912a, 438-441).

On page 22R of the *Zurich Notebook*, perhaps at the suggestion of Marcel Grossmann, Einstein wrote the Ricci tensor in terms of the Christoffel symbols $T_{il}$ and their derivatives. This way he obtained a fully covariant Ricci tensor in a form resulting from contraction of the Riemann tensor. Einstein divided the Ricci tensor into two parts – each of which separately transforms as a tensor under unimodular transformations – a tensor of second rank and a presumed gravitational tensor (Einstein 1912a, 451, 453). The presumed gravitational tensor was called by scholars the "November tensor":

$$(1a)\ T_{il}^x = \sum_{\kappa l} \frac{\partial}{\partial x_\kappa} \begin{Bmatrix} il \\ \kappa \end{Bmatrix} - \begin{Bmatrix} i\kappa \\ \lambda \end{Bmatrix} \begin{Bmatrix} \lambda l \\ \kappa \end{Bmatrix}.$$

where $\begin{Bmatrix} il \\ \kappa \end{Bmatrix}$ are the Christoffel symbols of the second kind, see equation (5) below.

Setting the November tensor $T_{il}^x$ equal to the stress-energy tensor, multiplied by the gravitational constant, one arrives at the field equations of Einstein's first paper of November 4, 1915.

Einstein no longer required the harmonic coordinate condition, and he could impose the Hertz coordinate condition to eliminate all unwanted second-order derivative terms. He hoped he could extract the Newtonian limit from the November tensor $T_{il}^x$. The Hertz condition also ensured that the divergence of the linearized stress-energy tensor vanished, thus satisfying energy-momentum conservation law. Einstein then imposed the Hertz condition on the first part of the Ricci tensor, the tensor of second rank, and was able to recover an expression that could lead him to Newton's law of gravitation:

$(3a)$ $-\dfrac{1}{2}\sum \gamma_{\kappa\alpha}\dfrac{\partial^2 g_{il}}{\partial x_\alpha \partial x_\kappa} - \dfrac{1}{2}\sum \left(\dfrac{\partial \gamma_{\kappa\alpha}}{\partial x_l}\dfrac{\partial g_{i\alpha}}{\partial x_\kappa} + \dfrac{\partial \gamma_{\kappa\alpha}}{\partial x_i}\dfrac{\partial g_{l\alpha}}{\partial x_\kappa}\right).$

He now imposed the Hertz condition on the second part of the Ricci tensor, on $T_{il}^x$, and recovered an expression that could lead him to Newton's law of gravitation:

$(3b)$ $\sum \dfrac{1}{2}\gamma_{\alpha\beta}\dfrac{\partial^2 g_{il}}{\partial x_\alpha \partial x_\beta} - \dfrac{1}{4}\sum \gamma_{\alpha\kappa}\gamma_{\beta\lambda}\left(\dfrac{\partial g_{i\alpha}}{\partial x_\beta} - \dfrac{\partial g_{i\beta}}{\partial x_\alpha}\right)\left(\dfrac{\partial g_{l\kappa}}{\partial x_\lambda} - \dfrac{\partial g_{l\lambda}}{\partial x_\kappa}\right) +$

additional terms with products of first-order derivatives and Christoffel symbols of the first kind.

This expression eventually produces a Newton-Poisson equation as a first approximation.

On page 23L Einstein imposed a new coordinate condition to eliminate terms with unwanted second-order derivatives of the metric, and by which he hoped to extract the Newtonian limit from the November tensor $T_{il}^x$: the so-called "theta $\vartheta$ restriction" (see Section 2 for a detailed explanation on the $\vartheta$ restriction). He used the Hertz and $\vartheta$ conditions to eliminate various terms from $T_{il}^x$. The combination of the Hertz condition and the $\vartheta$ condition allowed Einstein to recover an expression from which the Newton-Poisson equation could again be obtained as a first approximation (Einstein 1912a, 454, 456):

$(3c)$ $\sum \gamma_{\kappa\alpha}\dfrac{\partial^2 g_{il}}{\partial x_\kappa \partial x_\alpha} + \sum \gamma_{\rho\alpha}\gamma_{k\beta}\dfrac{\partial g_{ik}}{\partial x_\alpha}\dfrac{\partial g_{l\rho}}{\partial x_\beta}.$

However, Einstein finally rejected both the truncated November tensor $T_{il}^x$ and equations (3b) and (3c) (Einstein 1912a, 450-454, 456). He then abandoned the Hertz condition, and wrote that it was unnecessary.

Einstein already accumulated coordinate conditions (the harmonic condition, the Hertz condition, and the $\vartheta$ condition) to eliminate the terms from his equations, and he extracted expressions of broad covariance from the Ricci tensor. He then truncated them by imposing additional conditions on the metric to obtain candidates for the left-hand side of the field equations that reduce to the Newtonian limit in the case of weak static fields. However, this model entangled the Newtonian limit and conservation of momentum-energy.[1]

On pages 24R-25R, Einstein therefore tried to extract yet another candidate for the left-hand side of the field equations. He did not extract these field equations from the Ricci tensor but established field equations while starting from the requirement of the conservation of momentum and energy. He still hoped to connect the new field equations he found through energy-momentum considerations to the November tensor $T_{il}^x$ of page 22R. However, he finally abandoned general covariance on the very next pages (26L and 26R) and derived the so-called

---

[1] This topic will be further explained in Section 2 devoted to interpretation of the *Zurich Notebook* by a group of scholars.

*Entwurf* field equations – which he also established using the same method, through energy-momentum considerations. These equations could be covariant with respect to linear transformations, and they satisfied both the Newtonian limit and conservation of momentum-energy (Einstein 1912a, 459, 461-469).

Three years later, in his first November 1915 paper on general relativity, Einstein returned to the November tensor $T_{il}^x$ and to the Hertz condition. He ended his paper by writing (Einstein 1915a, 786): "We can therefore set as a first-order approximation arbitrarily":

$$(2b) \sum_\beta \frac{\partial g^{\alpha\beta}}{\partial x_\beta} = 0.$$

This is the good old Hertz condition, equation (2a). Einstein used the Hertz condition to rewrite his field equations in the following form, the first term of equation (3b):

$$(3d) \frac{1}{2} \sum_\alpha \frac{\partial^2 g_{\mu\nu}}{\partial x_\alpha^2} = \kappa T_{\mu\nu},$$

from which one obtains the Newton-Poisson equation as a first approximation.

Einstein performed the following mathematical derivation (Einstein 1915a, 785-786):

Consider Einstein's field equations of November 4, 1915:

$$(4) \ R_{\mu\nu} = -\kappa T_{\mu\nu},$$

the November tensor $R_{\mu\nu}$ equal to the stress-energy tensor $T_{\mu\nu}$ multiplied by the gravitational constant $\kappa$.

Note that the stress-energy tensor $T_{\mu\nu}$ is a *covariant* tensor of the second-order and the $\mu\nu$ is placed as a subscript. $T^{\mu\nu}$ represents the components of the *contravariant* stress-energy tensor of the second-order. Hence, the $\mu\nu$ is placed as a superscript.

Einstein wrote his field equations (4) in full form:

$$(4a) \sum_\alpha \frac{\partial \Gamma_{\mu\nu}^\alpha}{\partial x_\alpha} + \sum_{\alpha\beta} \Gamma_{\mu\beta}^\alpha \Gamma_{\nu\alpha}^\beta = -\kappa T_{\mu\nu},$$

$\Gamma_{\mu\nu}^\alpha$ are the Christoffel symbols of the second kind:

(5) $\Gamma_{\mu\nu}^{\sigma} = -\left\{\begin{matrix}\mu\nu\\\sigma\end{matrix}\right\} = -\sum_{\alpha}g^{\sigma\alpha}\left[\begin{matrix}\mu\nu\\\alpha\end{matrix}\right] = -\frac{1}{2}\sum_{\alpha}g^{\sigma\alpha}\left(\frac{\partial g_{\mu\alpha}}{\partial x_{\nu}} + \frac{\partial g_{\nu\alpha}}{\partial x_{\mu}} - \frac{\partial g_{\mu\nu}}{\partial x_{\alpha}}\right),$

where $\left[\begin{matrix}\mu\nu\\\alpha\end{matrix}\right]$ are the Christoffel symbols of the first kind.

These represent the components of the gravitational field.

Einstein fully contracted the November tensor (the Ricci tensor) in his field equations (4a). He multiplied his field equations (4a) by $g^{\mu\nu}$ and summed over $\mu$ and $\nu$; and using the following equation:

(6) $\sum_{\tau}\left\{\begin{matrix}s\tau\\s\end{matrix}\right\} = \frac{1}{2}\sum_{\alpha s}g^{s\alpha}\left(\frac{\partial g_{s\alpha}}{\partial x_{\tau}} + \frac{\partial g_{\tau\alpha}}{\partial x_{s}} - \frac{\partial g_{s\tau}}{\partial x_{\alpha}}\right) = \frac{1}{2}\sum g^{s\alpha}\frac{\partial g_{s\alpha}}{\partial x_{\tau}} = \frac{\partial\left(\log\sqrt{-g}\right)}{\partial x_{\tau}},$

and after some rearranging, he obtained:

(7) $\sum_{\alpha\beta}\frac{\partial^{2}g^{\alpha\beta}}{\partial x_{\alpha}\partial x_{\beta}} - \sum_{\sigma\tau\alpha\beta}g^{\sigma\tau}\Gamma_{\sigma\beta}^{\alpha}\Gamma_{\tau\alpha}^{\beta} + \sum_{\alpha\beta}\frac{\partial}{\partial x_{\alpha}}\left(g^{\alpha\beta}\frac{\partial\log\sqrt{-g}}{\partial x_{\beta}}\right) = -\kappa\sum_{\sigma}T_{\sigma}^{\sigma} = -\kappa T.$

$T$ is the trace of the stress-energy tensor of matter.

In his 1914 summarizing long review article on his *Entwurf* theory, "The Formal Foundation of the General Theory of Relativity", Einstein showed that there is a relation between (5), the Christoffel symbols of the second kind, and the derivative of the determinant $g$ of the metric tensor $g_{\mu\nu}$. He first derived the expression for the derivative of the metric tensor, i.e. he differentiated the determinant of the metric tensor $|g_{\mu\nu}| = g$ with respect to $x_{\alpha}$:

(6a) $\frac{1}{g}\frac{\partial g}{\partial x_{\alpha}} = \sum g^{\mu\nu}\frac{\partial g_{\mu\nu}}{\partial x_{\alpha}} = \frac{\partial\left(\log\sqrt{-g}\right)}{\partial x_{\alpha}}.$

Einstein then contracted the Christoffel symbols of the second kind in equation (5) and obtained equation (6) (Einstein 1914, 1051).

He wrote the energy components of the gravitational field $t_{\sigma}^{\lambda}$, the components representing the energy and momentum stored in the gravitational field:

(8) $\kappa t_{\sigma}^{\nu} = \frac{1}{2}\delta_{\sigma}^{\nu}\sum_{\mu\nu\alpha\beta}g^{\mu\nu}\Gamma_{\mu\beta}^{\alpha}\Gamma_{\nu\alpha}^{\beta} - \sum_{\mu\nu\alpha}g^{\mu\nu}\Gamma_{\mu\sigma}^{\alpha}\Gamma_{\nu\alpha}^{\nu}.$

This is *a pseudo-tensor* which has tensorial character only under linear transformations.

$\delta_\sigma^\nu$ is the Kronecker delta. It is defined by:

$$\delta_\sigma^\nu = 1, \qquad \text{if } \nu = \sigma, \qquad \delta_\sigma^\nu = 0, \qquad \text{if } \nu \neq \sigma.$$

The energy-momentum balance for matter in a gravitational field reads:

$$(8a) \quad \sum_\alpha \frac{\partial T_\sigma^\alpha}{\partial x_\alpha} = -\sum_{\alpha\beta} \Gamma_{\sigma\beta}^\alpha T_\alpha^\beta.$$

$T_\sigma^\alpha$ is the stress-energy tensor of matter, a mixed tensor and it transforms like a contravariant tensor with respect to the index $\alpha$ and like a covariant tensor with respect to the index $\sigma$. Hence, the $\alpha$ is placed as a superscript while the $\sigma$ is placed as a subscript.

Einstein now contracted (4a), i.e. he multiplied (4a) by $g^{\nu\lambda}$ and summed over $\nu$:

$$(4b) \quad \sum_{\alpha\nu} \frac{\partial}{\partial x_\alpha} \left( g^{\nu\lambda} \Gamma_{\mu\nu}^\alpha \right) - \sum_{\alpha\beta\nu} g^{\nu\beta} \Gamma_{\nu\mu}^\alpha \Gamma_{\beta\alpha}^\lambda = -\kappa T_\mu^\tau.$$

The second term on the left-hand side of (4b) is the second term on the right-hand side of (8) $t_\sigma^\nu$, and so Einstein could combine between the two and obtain:

$$(9) \quad \sum_{\alpha\nu} \frac{\partial}{\partial x_\alpha} \left( g^{\nu\lambda} \Gamma_{\mu\nu}^\alpha \right) - \frac{1}{2} \delta_\mu^\lambda \sum_{\mu\nu\alpha\beta} g^{\mu\nu} \Gamma_{\mu\beta}^\alpha \Gamma_{\nu\alpha}^\beta = -\kappa \left( T_\mu^\tau + t_\mu^\lambda \right).$$

Thus according to conservation of energy-momentum:[2]

$$(10) \quad \sum_\lambda \frac{\partial}{\partial x_\lambda} \left( T_\sigma^\lambda + t_\sigma^\lambda \right) = 0,$$

the divergence of the left-hand side of (9) equals zero:

$$(11) \quad \frac{\partial}{\partial x_\mu} \left[ \sum_{\alpha\beta} \frac{\partial^2 g^{\alpha\beta}}{\partial x_\alpha \partial x_\beta} - \sum_{\sigma\tau\alpha\beta} g^{\sigma\tau} \Gamma_{\sigma\beta}^\alpha \Gamma_{\tau\alpha}^\beta \right] = 0,$$

---

[2] Einstein first demonstrated that equation (10) follows from his field equations (4a), see equation (34a), (34b) and (8h) in Section 3.2. The above derivation [equations (7)–(9)] came afterwards.

and the first two left-hand side terms of equation (7) vanish:

$$(11a) \quad \sum_{\alpha\beta} \frac{\partial^2 g^{\alpha\beta}}{\partial x_\alpha \, \partial x_\beta} - \sum_{\sigma\tau\alpha\beta} g^{\sigma\tau} \Gamma^\alpha_{\sigma\beta} \Gamma^\beta_{\tau\alpha} = 0.$$

This equation "tells us that the coordinate system must be adapted to the manifold" (Einstein 1915, 785).

Einstein then wrote that a first approximation of equation (11a) is:

$$(11b) \quad \sum_{\alpha\beta} \frac{\partial^2 g^{\alpha\beta}}{\partial x_\alpha \, \partial x_\beta} = 0,$$

and that this equation does not yet fix the coordinate system because it would require four equations (Einstein 1915, 786): "We can therefore set as a first-order approximation arbitrarily" equation (2b) the Hertz coordinate condition.

Einstein used the Hertz coordinate condition to rewrite his field equations (4a) in the form (3d).

However, the relation between (5) and (6a) caused a problem because according to (11a) and (7):

$$(7a) \quad \sum_{\alpha\beta} \frac{\partial}{\partial x_\alpha} \left( g^{\alpha\beta} \frac{\partial \log\sqrt{-g}}{\partial x_\beta} \right) = -\kappa T.$$

This equation says that it is impossible to choose a coordinate system in which $\sqrt{-g} = 1$, for if we choose such a coordinate system, then $\log 1 = 0$ and the trace $\sum_\sigma T^\sigma_\sigma = T$ of the stress-energy tensor vanishes (Einstein 1915a, 785).

In fact, Einstein solved this problem when he formulated the final form of the field equations, which was announced on November 25, 1915. I very briefly trace the solution. In the November 25 paper, Einstein fully contracted the Ricci tensor in his field equations (Einstein 1915c, 846-847):

$$(4c) \quad G_{im} = -\kappa \left( T_{im} - \frac{1}{2} g_{im} T \right),$$

and he got the equation:

$$(7b) \quad \sum_{\alpha\beta} \frac{\partial^2 g^{\alpha\beta}}{\partial x_\alpha \, \partial x_\beta} - \sum_{\sigma\tau\alpha\beta} g^{\sigma\tau} \Gamma^\alpha_{\sigma\beta} \Gamma^\beta_{\tau\alpha} + \sum_{\alpha\beta} \frac{\partial}{\partial x_\alpha} \left( g^{\alpha\beta} \frac{\partial \log\sqrt{-g}}{\partial x_\beta} \right) = \kappa T.$$

Inserting the trace of the stress-energy pseudo-tensor of the gravitational field:

$$\kappa t = \sum_{\sigma\tau\alpha\beta} g^{\sigma\tau} \Gamma^{\alpha}_{\sigma\beta} \Gamma^{\beta}_{\tau\alpha}$$

into equations (7b) leads to:

$$(7c) \sum_{\alpha\beta} \frac{\partial^2 g^{\alpha\beta}}{\partial x_{\alpha} \, \partial x_{\beta}} - \kappa t + \sum_{\alpha\beta} \frac{\partial}{\partial x_{\alpha}} \left( g^{\alpha\beta} \frac{\partial \log\sqrt{-g}}{\partial x_{\beta}} \right) = \kappa T.$$

Instead of (11) and (11a), Einstein obtained:

$$(11c) \frac{\partial}{\partial x_{\mu}} \left[ \sum_{\alpha\beta} \frac{\partial^2 g^{\alpha\beta}}{\partial x_{\alpha} \, \partial x_{\beta}} - \kappa(T+t) \right] = 0 \text{ and:}$$

$$(11d) \sum_{\alpha\beta} \frac{\partial^2 g^{\alpha\beta}}{\partial x_{\alpha} \, \partial x_{\beta}} - \kappa(T+t) = 0.$$

The term $t$, representing the trace of the stress-energy pseudo-tensor of the gravitational field, occurs on an equal footing with $T$, the trace of the stress-energy tensor of matter. This, said Einstein, was not the case in his first November 4, 1915 paper.

According to (11c) and (11d):

$$(7d) \sum_{\alpha\beta} \frac{\partial}{\partial x_{\alpha}} \left( g^{\alpha\beta} \frac{\partial \lg\sqrt{-g}}{\partial x_{\beta}} \right) = 0,$$

Why did Einstein reject the November tensor $T^{x}_{il}$ in 1912-1913, only to come back to it three years later? On November 18, 1915, Einstein wrote to David Hilbert that: "The difficulty was not in finding generally covariant equations for the [metric tensor]; for this is easily achieved with the aid of Riemann's tensor. Rather, it was hard to recognize that these equations are… a simple and natural generalization of Newton's law. It has just been in the last few weeks that I succeeded in this… whereas 3 years ago with my friend Grossmann I had already taken into consideration the only possible generally covariant equations, which have now been shown to be the correct ones. We had only distanced ourselves with heavy heart from it, because it seemed to me that the physical discussion had shown their incompatibility with Newton's law" (Einstein to Hilbert, November 18, 1915, *CPAE 8*, Doc. 148).

Let us pose the following questions: Einstein said that the difficulty was not in finding equation (1a), the November tensor $T^{x}_{il}$. 1) Why then did he reject this tensor in 1912-1913, only to come back to it in November 1915? Einstein explained that he found it difficult to recognize that the November tensor $T_{il}^x$ reduces to the Newtonian limit. 2) Why was it hard for Einstein to recognize that the generally covariant equations are a simple and natural generalization of Newton's law of gravitation? 3) Why did it take him three years to arrive at the realization that the November tensor $T_{il}^x$ is not incompatible with Newton's law? I will consider each of these three separate questions in turn.

My analysis of Einstein's *Zurich Notebook* is based on a critical interpretation of the notebook guided by a group of scholars. This work can only be summarized here. However, it goes without saying that the interpretation of Einstein's papers and *Zurich Notebook* which is provided in this paper reflect my own understanding of Einstein's works and of scholars' papers.

In Section 2 I present the main aspects, which are relevant to my discussion. In Section 3 I discuss several conjectures and possible answers to the three questions. Several scholars have reached different conclusions from Einstein's *Zurich Notebook*, calculations and texts. It seems that no study today has been able to put to rest the question: Why did Einstein reject the November tensor $T_{il}^x$, equation (1a), in 1912-1913, only to come back to it three years later? Since the case is not yet resolved, in Section 4 I advance a combined conjecture in light of the available evidence.

## 2 Interpretation of the *Zurich Notebook*

In the course of preparing the editorial project of the *Collected Papers of Albert Einstein* John Stachel first realized the significance of the *Zurich Notebook* for the reconstruction of the genesis of general relativity (Renn and Sauer 1999, 89). A group of leading scholars (Michel Janssen, John Norton, Jürgen Renn, Tilman Sauer and John Stachel) have been actively doing research on the *Zurich Notebook* and related material for many years. They published their results in the monumental 4-volume book, *The Genesis of General Relativity*, and the results are briefly summarized below (Einstein 1912a, 451, 453-469; Janssen, Renn, Sauer, Norton and Stachel, 2007, 607-608, 627, 645-656, 661-679, 681-683, 687-688, 695-696, 704). This section describes the common ground shared by the authors of the book, *The Genesis of General Relativity*.

Einstein's calculations and comments in the *Zurich Notebook* provide strong evidence in support of the conjecture that he imposed the Hertz condition [equation (2a)] on the November tensor $T_{il}^x$, equation (1a), to restrict the covariance of the tensor $T_{il}^x$. He thus used the Hertz condition as a "Hertz restriction". Jürgen Renn first made the distinction between coordinate conditions and what he has called "coordinate restrictions" (Janssen and Renn 2007, 886). Historians of general relativity favor this nomenclature with which they distinguish restricting covariance from recovering the Newtonian limit.

More specifically, a coordinate condition (in the modern sense) is used to recover the Newtonian limit from the gravitational field equations because the Poisson equation is not generally covariant; it is only covariant under Galilean transformations. Therefore, one imposes conditions on the field equations of broader covariance if one wants to recover the Newtonian limit.

Coordinate conditions do not restrict the covariance of the field equations; they are only used to choose coordinate systems in the context of obtaining the Newtonian limit.

A coordinate restriction is imposed on the field equations. The Hertz condition (2a) eliminates various terms from the November tensor $T_{il}^x$ by restricting the admissible coordinate systems to systems which satisfy this coordinate restriction. Consequently, Einstein attempted at eliminating unwanted second-order derivative terms from the tensor $T_{il}^x$ with the help of the Hertz restriction (2a) and under the unimodular restriction. Thus, in the *Zurich Notebook* Einstein used coordinate restrictions, which restricted the covariance of the field equations (Räz 2016).

The Hertz condition served two purposes:

1) It eliminated unwanted terms with second-order derivatives of the metric from the tensor $T_{il}^x$.

2) It ensured that the divergence of the matter stress-energy tensor vanished in a weak-field approximation.

At the bottom of page 22R, Einstein extracted a candidate for the left-hand side of the field equations from the November tensor $T_{il}^x$ with the Hertz restriction. The field equations based on this reduced tensor $T_{il}^x$ are covariant under those unimodular transformations (unimodular because the determinant of the metric is set equal to unity in all calculations) that preserve the Hertz restriction (2a).

Scholars have called the expression in equation (2a):

$$(2c) \quad \sum \gamma_{\kappa\kappa} \frac{\partial g_{i\kappa}}{\partial x_\kappa},$$

the "Hertz expression". In equation (2a) the Hertz expression is set equal to zero.

On page 22L Einstein checked whether the Hertz restriction (2a) can transform as a tensor under restriction to unimodular transformations: Given a metric field satisfying the Hertz restriction in some (unprimed) coordinate system, what are the unimodular coordinate transformations such that the Hertz restriction will be satisfied in the new (primed) coordinate system as well? Einstein had found that the Hertz restriction does not allow a transformation to uniformly accelerated frames of reference in Minkowski spacetime. He nonetheless continued to use the Hertz restriction to eliminate unwanted second-order derivative terms from the November tensor $T_{il}^x$.

The candidate extracted from the November tensor $T_{il}^x$ contains numerous terms with products of first-order derivatives of the metric. Most of these terms come from the product of the Christoffel symbols in the second term of the tensor $T_{il}^x$. However, Einstein realized that he needed an additional coordinate restriction to eliminate two of the three first-order derivative terms in every Christoffel symbol, that is to say, terms quadratic in first-order derivatives of the metric tensor from $T_{il}^x$; the Hertz restriction would still be needed to eliminate terms with unwanted second-order derivatives.

On page 23L Einstein thus returned to the expression for the November tensor $T_{il}^x$ in terms of the Christoffel symbols. He imposed a new coordinate restriction, the so-called "$\vartheta$ coordinate restriction", on the tensor $T_{il}^x$, with which he could eliminate terms quadratic in first derivatives of the metric from $T_{il}^x$; and by which he hoped to extract the Newtonian limit from this tensor. Einstein required that the $\vartheta$ restriction would limit the allowed coordinate transformations to $\vartheta$ transformations under which the following quantity (called by scholars, "$\vartheta$ expression"):

$$(12) \ \ \vartheta_{ik\lambda} = \frac{1}{2}\left(\frac{\partial g_{ik}}{\partial x_\lambda} + \frac{\partial g_{k\lambda}}{\partial x_i} + \frac{\partial g_{\lambda i}}{\partial x_k}\right),$$

transformed as a tensor, and by which Einstein could eliminate unwanted terms from the November tensor $T_{il}^x$.

Scholars explain that the convention in the *Zurich Notebook* is the following: Einstein used Greek and Latin characters to denote covariant and contravariant quantities, respectively. Consequently, $\vartheta_{ik\lambda}$ represents a covariant version of the $\vartheta$ expression.

Einstein obtained a reduced November tensor $T_{il}^x$ and candidate field equations that were invariant under $\vartheta$ transformations. He used the $\vartheta$ expression (12) to rewrite the Christoffel symbols of the first and second kind, and obtained:

$$(13) \ \begin{bmatrix} il \\ k \end{bmatrix} = \vartheta_{ilk} - \frac{\partial g_{il}}{\partial x_k}, \ \ \ \begin{Bmatrix} il \\ k \end{Bmatrix} = \gamma_{k\alpha}\left(\vartheta_{il\alpha} - \frac{\partial g_{il}}{\partial x_\alpha}\right), \ \ \ \ \ \text{respectively.}$$

He then substituted expression (13) into the November tensor $T_{il}^x$, equation (1a), and arrived at:

$$(13a) \ T_{il}^x = \sum_{\kappa l} \frac{\partial}{\partial x_\kappa} \gamma_{k\alpha}\left(\vartheta_{il\alpha} - \frac{\partial g_{il}}{\partial x_\alpha}\right) - \gamma_{\lambda\alpha}\gamma_{k\beta}\left(\vartheta_{ik\alpha} - \frac{\partial g_{ik}}{\partial x_\alpha}\right)\left(\vartheta_{i\lambda\beta} - \frac{\partial g_{l\lambda}}{\partial x_\beta}\right).$$

If we set: $\vartheta_{il\alpha} = 0$, then the above equation becomes:

$$(14) \sum_{\kappa l} \frac{\partial}{\partial x_\kappa} \gamma_{k\alpha}\left(\frac{\partial g_{il}}{\partial x_\alpha}\right) + \sum \gamma_{\rho\alpha}\gamma_{k\beta}\frac{\partial g_{ik}}{\partial x_\alpha}\frac{\partial g_{l\rho}}{\partial x_\beta}.$$

Finally, the combination of the Hertz restriction and the $\vartheta$ restriction allowed Einstein to extract from the November tensor $T_{il}^x$ a candidate for the left-hand side of the field equations. He obtained a reduced November tensor $T_{il}^x$ and was able to recover an expression from the reduced $T_{il}^x$ that could lead him to Newton's law of gravitation [equation (3c)]. The second term on the right-hand side of (14) is the second term on the right-hand side of equation (3c).

It seems likely that at this point Einstein may have noticed that $\gamma_{k\alpha}\left(\frac{\partial g_{il}}{\partial x_\alpha}\right)$ in the first term on the left hand side of (14) was nothing but the truncated Christoffel symbols. He could just as well insert the truncated Christoffel symbols into the November tensor $T_{il}^x$ and thereby obtain equation (14). As it occurs in Einstein's considerations in the notebook, he came to realize that with the $\vartheta$ restriction he did not need the Hertz restriction, equation (2a), anymore because the $\vartheta$ restriction not only eliminates additional terms but it also takes care of the unwanted terms of second-order derivatives in the November tensor $T_{il}^x$ that he had eliminated earlier with the Hertz restriction. He therefore abandoned the Hertz restriction and wrote that it was unnecessary. Just as he had done with the Hertz restriction, he then checked whether the $\vartheta$ expression (12) transforms as a tensor.

However, Einstein rejected the truncated November tensor $T_{il}^x$ and the expression that could lead him to the Poisson-Newton equation. He found that the $\vartheta$ expression (12) vanishes for a so-called rotation metric, which the scholars called the "$\vartheta$ rotation metric", or in short, the "$\vartheta$ metric". Having inverted this metric, Einstein realized that its components are exactly those of the Minkowski metric in rotating coordinates (the rotation metric). For this reason, Einstein attempted to demonstrate that since the inverse of the $\vartheta$ metric is equivalent to the Minkowski rotation metric, the rotation metric is definitely a solution of expression (12), i.e. of equation (12) below. However, this approach did not work. Let us examine this in more detail.

On pages 42L and 42R of the *Zurich Notebook*, Einstein tried to find whether the $\vartheta$ expression (12) satisfied the Minkowski metric in rotating coordinates (the rotation metric). The group of Einstein scholars conclude: The $\vartheta$ expression (12) was not a coordinate condition because Einstein thought he needed to show that the $\vartheta$ expression (12) vanished for the Minkowski metric in rotating coordinates. They explain that the precise temporal order of the calculations on pages 23L–24L at one end of the *Zurich Notebook* and on pages 42L–43L at the other remains unclear. Einstein might have switched back and forth between these two sets of pages.

The *covariant* second-order components of the rotation metric for counterclockwise rotation around the *z*-axis with angular velocity ω are:

(15)
$$
\begin{matrix}
-1 & 0 & \omega y \\
0 & -1 & -\omega x \\
\omega y & -\omega x & 1 - \omega^2(x^2 + y^2)
\end{matrix}
$$

For clockwise rotation around the *z*-axis with angular velocity ω, the *contravariant* rotation metric is given by:

(16)
$$
\begin{matrix}
-1 + \omega^2 y^2 & -\omega^2 xy & \omega y \\
-\omega^2 xy & -1 + \omega^2 x^2 & -\omega x \\
\omega y & -\omega x & 1
\end{matrix}
$$

(Einstein wrote the components of the metric suppressing one spatial dimension).

Einstein wrote the components of the metric tensor, and then tried to find the most general solution of the $\vartheta$ expression (12) for this metric field:

$$(12a)\ \vartheta_{ik\lambda} = \frac{1}{2}\left(\frac{\partial g_{ik}}{\partial x_\lambda} + \frac{\partial g_{k\lambda}}{\partial x_i} + \frac{\partial g_{\lambda i}}{\partial x_k}\right) = 0.$$

On page 42R, he imposed the condition $\sqrt{-g} = 1$ (a restriction to unimodular transformations).

He found that equation (12a) vanishes for the following solution, a so-called rotation metric, which scholars called the "$\vartheta$ rotation metric" or "$\vartheta$ metric" (in its *covariant* form):

$$(17)\quad \begin{matrix} -(1 + \alpha x_2^2) & \alpha x_1 x_2 & \beta x_2 \\ \alpha x_1 x_2 & -(1 + \alpha x_1^2) & -\beta x_1 \\ \beta x_2 & -\beta x_1 & 1 \end{matrix}$$

where $\alpha$ and $\beta$ are integration constants.

Therefore (17) is the solution of (12a). The determinant of the above *covariant* $\vartheta$ metric is not 1. Einstein added the determinant 1 condition, $\alpha = -\beta^2$; he then inverted the above covariant $\vartheta$ metric and obtained a *contravariant* $\vartheta$ metric:

$$(18)\quad \begin{matrix} -1 & 0 & \beta x_2 \\ 0 & -1 & -\beta x_1 \\ \beta x_2 & -\beta x_1 & 1 + \alpha(x_1^2 + x_2^2) \end{matrix}$$

whose components are the components of the *covariant* Minkowski metric in rotating coordinates (15). When the constant $\beta$ in the above metric is set equal to the angular velocity $\omega$, and since $\alpha = -\beta^2$, then $\alpha = -\omega^2$, we obtain the covariant second-order Minkowski metric in rotating coordinates (15).

There was a remarkable coincidence between (18) and (15) which indicated a degree of similarity of the two. Perhaps, Einstein said to himself: I think indeed that since the *covariant* $\vartheta$ rotation metric (17) of which I have shown that (18) is the inverse, is a solution of $\vartheta_{ik\lambda} = 0$, then the *covariant* Minkowski metric in rotating coordinates (15) is definitely a solution of $\vartheta_{ik\lambda} = 0$. However, this most likely approach did not work and it was puzzling in light of the similarity of (15) and the inverse of the $\vartheta$ metric, equation (18). The final result was: $\vartheta_{ik\lambda} = 0$ and the $\vartheta$ expression did not allow the Minkowski metric in rotating coordinates (15), but they did allow the $\vartheta$ metric (17). After all, the $\vartheta$ metric was not an ordinary rotation metric but the inverse of the $\vartheta$ metric, equation (18), definitely gives the impression of a rotation metric. This is true but Einstein was stuck to the $\vartheta$ metric.

Such was the story of the $\vartheta$ metric. Einstein thought that the $\vartheta$ metric (17) might be interpreted in terms of inertial forces in rotating frames of reference. Scholars discovered that on pages 42R and 43LA, Einstein wanted to check whether the components of the $\vartheta$ rotation metric (17) and their derivatives could be given the same sort of physical meaning in terms of inertial forces in rotating frames of reference as the components of the Minkowski metric (15) and their derivatives in rotating coordinates. Einstein considered the motion of a material point in the $\vartheta$ rotation metric and inserted the $\vartheta$ metric in its *covariant* form (17) into a Lagrangian [see equation (25b) of Section 3.1]. He then wrote the variational principle and derived the equations of motion as the Euler-Lagrange equations of a particle moving in a metric field:

(19) $\dfrac{d}{dt}\left(\dfrac{\partial L}{\partial \dot{x}}\right) - \dfrac{\partial L}{\partial x} = 0.$

The calculation did not produce a satisfactory result.

Scholars note that on page 24L, Einstein tried to interpret the components of the *contravariant $\vartheta$* metric (18) in terms of inertial forces in rotating frames of reference, just as one would do for the ordinary Minkowski metric in rotating coordinates (15). He thus checked whether the force on a particle at rest in the gravitational field described by the metric (18) could be interpreted as the centrifugal force on a particle at rest in a rotating frame of reference.

In as few words as I may, I will explain Einstein's calculation step by step: First, Einstein wrote an equation on page 05R that expresses the energy-momentum balance for matter in a gravitational field. This equation states that the covariant divergence of the contravariant stress-energy tensor $T_{\mu\nu}$ for pressureless dust (for matter) vanishes. It is the sum of two terms: the first term is the ordinary divergence of the stress-energy tensor for pressureless dust and the second term is an expression, which again contains this tensor and can be interpreted as the gravitational force on a particle. This equation with a determinant $g$ equal to unity (with $\sqrt{g} = 1$) in linear approximation is the following:

(8b) $\displaystyle\sum \dfrac{\partial}{\partial x_n}\left(g_{\mu\nu}T_{\nu n}\right) - \dfrac{1}{2}\dfrac{\partial g_{\mu\nu}}{\partial x_m}T_{\mu\nu} = 0.$

Brian Pitts has suggested that Einstein used a particle matter action principle (which leads to a geodesic equation) to obtain the law that expresses the energy-momentum balance for matter in a gravitational field (Pitts 2016, 59-60). Indeed, in his *Zurich Notebook* Einstein obtained the energy-momentum balance for matter in a gravitational field and the second term on the left-hand side of this equation by starting from an action principle, equation (25a) (see Section 3.1) and the equations of motion of a material particle in a metric field. The action integral is the proper length of the particle's world-line. Einstein wrote down the line element, equation (28) (see Section 3.2), and used it to define the Lagrangian, equation (25b) (see Section 3.1). He then wrote down (19), and the *x*-component of the momentum density of pressureless dust. An expression for the energy density of pressureless dust was not written down explicitly in the notebook but both expressions give the components of the contravariant stress-energy tensor for pressureless dust in equation (1). Taking into consideration equation (28) and equation (25b)

Einstein generalized this term to the following expression (Einstein 1912, 383, 385; Janssen, Renn, Sauer, Norton and Stachel, 2007, 516-520):

$$(19a) \quad -\frac{1}{2} \sum \sqrt{g} \, \frac{\partial g_{\mu\nu}}{\partial x_m} T_{\mu\nu}^b,$$

which represents the force density acting on the cloud of dust due to a gravitational field. This term would reappear as the gradient of the metric representing the components of the gravitational field in Einstein's 1914 *Entwurf* theory.

Subsequently Einstein inserted an expression for the tensor of momentum and energy and (19a) into (19) that he started from and arrived at a candidate equation for the law that expresses the energy-momentum balance for matter in a gravitational field, equation (8b):

$$\sum_{\nu n} \frac{\partial}{\partial x_n} \left( \sqrt{g} \, g_{m\nu} T_{\nu n} \right) - \frac{1}{2} \sum_{\mu\nu} \sqrt{g} \, \frac{\partial g_{\mu\nu}}{\partial x_m} T_{\mu\nu} = 0.$$

Second, on page 43LB, Einstein demonstrated that the equation of motion for one particle can be obtained by integrating this equation with $T_{\mu\nu}$ given by equation (1), over the volume of the particle. He could thus look upon (8b) as giving the equation of motion of a particle in a metric field with $\sqrt{g} = 1$. In particular, he could look upon the second term on the left-hand side of this equation as giving the gravitational force on a particle at rest in a metric field $g_{\mu\nu}$, i.e. the gravitational force density experienced by a particle at rest in the metric field under consideration:

$$(8c) \quad \frac{1}{2} \frac{\partial g_{\mu\nu}}{\partial x_m} T_{\mu\nu}.$$

Einstein thus considered a particle at rest with respect to a rotating coordinate system in Minkowski spacetime. In that case, the contravariant stress-energy tensor for pressureless dust $T_{\mu\nu}$ in equation (1) reduces to: diag (0, 0, 0, $\rho_0/g_{44}$). Inserting this equation and the covariant Minkowski rotation metric $g_{\mu\nu}$, equation (15), into the $m = 1$ component of (8c), and when terms of higher order in the angular frequency of the rotating coordinate system $\omega^4$ are neglected, we obtain the $x$-component of the centrifugal force in ordinary Newtonian theory:

$$\frac{\rho_0}{g_{44}} \omega^2 x = \rho_0 \omega^2 x + \cancel{O(\omega^4)}.$$

Since the contravariant $\vartheta$ metric (18) has the same form as the Minkowski metric in rotating coordinates (15), Einstein wanted to obtain an analogous result: He wanted to check whether one could insert equation (18) into the second term of the energy-momentum balance between matter and gravitational field and interpret this term as the centrifugal force in ordinary Newtonian theory. For this purpose, he wrote the contravariant stress-energy tensor in terms of the covariant stress-energy tensor $\Theta_{\alpha\beta}$. He substituted the resulting expression into equation (8b). He obtained the second term, the gravitational force density experienced by a particle at rest in a metric field $\gamma_{\alpha\beta}$:

(8d) $\dfrac{1}{2}\dfrac{\partial \gamma_{\alpha\beta}}{\partial x_m}\Theta_{\alpha\beta}$.

Expression (8d) would give the same result for a particle at rest in the field of the $\vartheta$ metric (18) as expression (8c) for a particle at rest in (15), if the components of $\Theta_{\alpha\beta}$ in the former case were equal to those of $T_{\mu\nu} = $ diag $(0, 0, 0, \rho_0/g_{44})$ in the latter. Einstein thus inserted the following two equations into the $m = 1$ component of (8d): the simple expression, diag $(0, 0, 0, \rho_0)$ for $\Theta_{\alpha\beta}$ and the derivative of the components of the $\vartheta$ metric in its contravariant form (18) with respect to $x_1$: $\dfrac{d\gamma_{\alpha\beta}}{dx_1}$. He obtained:

$\dfrac{1}{2}\dfrac{\partial \gamma_{44}}{\partial x_1}\Theta_{44} = \rho_0 \alpha x_1,$

which is equal to the centrifugal force in Newtonian theory if the identification $\alpha = \omega^2$ is made.

What more could Einstein ask for? He found, as he had expected, that the $\vartheta$ metric (18) can thus be interpreted in terms of centrifugal forces just as the rotation metric (15). But no! Einstein still had a big little problem here. As if Einstein had not trouble enough dealing with the $\vartheta$ metric, the covariant stress-energy tensor was not diag $(0, 0, 0, \rho_0)$, it was much more complicated. In the case of a material point at rest with respect to the $\vartheta$ metric (18) we thus do not obtain this simple elegant result. Additional terms prevent us from obtaining the centrifugal force. Scholars have concluded that it is not entirely clear what conclusion Einstein drew from this calculation.

On page 43LA, Einstein then tried to modify the $\vartheta$ expression (12), rather than the original $\vartheta$ metric as it appeared until then and he wrote a contravariant version of (12) $t_{\lambda\alpha\kappa}$. Einstein then tried to find the most general solution of $t_{\lambda\alpha\kappa}$:

(12b) $t_{\lambda\alpha k} = \dfrac{1}{2}\left(\dfrac{\partial \gamma_{\lambda\alpha}}{\partial x_\kappa} + \dfrac{\partial \gamma_{\alpha k}}{\partial x_\lambda} + \dfrac{\partial \gamma_{\kappa\lambda}}{\partial x_\alpha}\right) = 0.$

He expected that this modified $\vartheta$ expression would vanish for the Minkowski metric in rotating coordinates (15). Since the inverse of the $\vartheta$ metric (17) is (18) and the inverse of (12a) is (12b), then (15) must be a solution of (12b). However, Einstein seems to have realized that the latter expression (12b) was not mathematically correct because of the intermingling of covariant and contravariant elements. His additional attempt at reconciling the $\vartheta$ expression with rotation had failed. Nevertheless, Einstein checked whether the modified $\vartheta$ expression could be used to eliminate terms with unwanted second-order derivatives of the metric from the November tensor $T_{il}^{x}$. Of course he found that, unlike the original $\vartheta$ expression (12), it could not, and he finally abandoned the attempt to modify the $\vartheta$ expression.

After a few failed attempts on pages 24L, 42L and 43L to come to terms with the $\vartheta$ metric, Einstein gave up the $\vartheta$ coordinate restriction, though not yet the idea of extracting field equations from the November tensor $T_{il}^x$. In the end however, he was forced to give up the $\vartheta$ restriction with which he constructed a candidate for the left-hand side of the field equations, because the $\vartheta$ expression conflicted with the Minkowski metric in uniformly rotating coordinates; it did not allow transformation to rotating frames in Minkowski spacetime. Einstein first attempted to resolve this problem in various ways but he finally gave up the truncated November tensor $T_{il}^x$.

I am now going to simplify Einstein's complicated derivations and scholars' lengthy reconstruction of his derivations. On page 24R Einstein tried to extract yet another candidate for the left-hand side of the field equations. He did not extract these field equations from the Ricci tensor but established field equations while starting from the requirement of the conservation of momentum and energy. In line with the Newton-Poisson equation, Einstein presumed that his new field equations would be of second-order. However, he was unable to find a candidate expression for the left hand side of the field equations that was a generalization of the Poisson equation and that proved to be a tensor with respect to arbitrary transformations. The equations could be covariant with respect to unimodular linear transformations, and they satisfied both the Newtonian limit and conservation of momentum-energy. Scholars conjecture that it is possible that Einstein set $\sqrt{-g} = 1$ to facilitate comparison of the result of his calculations with the November tensor $T_{il}^x$, which is a tensor only under unimodular transformations.

Einstein checked his new candidate using the contravariant rotation metric (16). He thought that his expression vanished for the rotation metric, a necessary condition for the rotation metric to be a solution of the vacuum field equations (in the absence of matter). The expression vanishing for the rotation metric may have signified to Einstein that the new field equations satisfied the relativity principle and the equivalence principle. He came to believe that his expression vanished for the rotation metric because of a few mistakes. He wrote the contravariant rotation metric in the following form:

$$(16a) \quad \begin{matrix} -1 + \omega^2 y^2 & -\omega xy & \omega y \\ -\omega xy & -1 + \omega^2 x^2 & -\omega x \\ \omega y & -\omega x & 1 \end{matrix}$$

instead of in the correct form (16). He may have somewhat mixed between the $\vartheta$ metric in its covariant form (17) and the Minkowski metric in rotating coordinates in contravariant form (16). He thus wrote ωxy instead of ω²xy.

Just then Einstein realized that the rotation metric was not a solution of his new field equations. Rather than find new field equations that do allow the rotation metric as a solution, on pages 25L and 25R Einstein tried to extract the new field equations of page 24R from the November tensor $T_{il}^x$. He almost succeeded in doing this but several unwanted terms thwarted his mission. Scholars conjecture that it was then that he tried to find a new modification of the $\vartheta$ restriction with the help of which he could eliminate all unwanted terms and derive the new field equations of page 24R from the November tensor $T_{il}^{x}$. Einstein therefore returned to the grand mission of pages 24L, 42L and 43L. He checked whether the new $\vartheta$ expression allows transformations to rotating frames of reference in Minkowski spacetime, and he returned to the good old $\vartheta$ metric and to the derivations of the above pages. There were certainly too many of $\vartheta$ elements in the air. Of course, the new $\vartheta$ expression did not vanish for the rotation metric, which is probably why Einstein abandoned the new $\vartheta$ restriction. Finally, the November tensor $T_{il}^{x}$ failed to yield the new field equations and Einstein failed in his quest for the "Holy Grail". He thus rejected his efforts of recovering his new field equations found through energy-momentum considerations from the tensor $T_{il}^{x}$, and he abandoned the latter. Einstein then wrote in the lower left corner of page 25R the word, "impossible".

And so ended that journey to save the November tensor. Not long after, on the very next pages (26L and 26R), Einstein presented the *Entwurf* field equations, which he also established using the same method, through energy-momentum considerations. Under the title, "System of Equations for Matter", Einstein derived gravitational field equations of limited covariance that were not derived from the Riemann tensor. He included the factor $\sqrt{-g}$ that was omitted on page 24R in the derivation of the *Entwurf* field equations. Perhaps he was no longer interested in trying to recover his new *Entwurf* field equations from the November tensor $T_{il}^{x}$. Good-bye November tensor. These equations spread over two facing pages, 26L and 26R, and are displayed with a neatness and order rare among the other pages of the notebook, suggesting that they were transcribed from another place (probably from Einstein's and Grossmann's joint 1913 *Entwurf* paper) after the result was known. Einstein ended his gravitation calculations on page 26R, with the left-hand side of the *Entwurf* gravitation tensor.

Einstein's collaboration with Grossmann finally took them away from general covariance and led them to a joint paper "Entwurf einer verallgemeinerten Relativitätstheorie und einer Theorie der Gravitation" (Outline of a Generalized Theory of Relativity and of a Theory of Gravitation) that was published before the end of June 1913 (Einstein and Grossmann 1913). Einstein wrote the physical part of the *Entwurf* paper and Grossmann wrote the mathematical part.

### 3. Competing Conjectures on Einstein's Rejection of the November Tensor

The only way we can study Einstein's route to general relativity is through primary sources and multiple perspectives of various historians. It is evident that several answers to the three questions posed at the end of Section 1 would be vital to the understanding of Einstein's rejection of the November tensor $T_{il}^{x}$, equation (1a), in 1912-1913. The answers of scholars I present below remain my interpretation of their conjectures. I hope this interpretation is as rich and creative as are the conjectures themselves. I start with John Stachel's conjecture:

### 3.1. John Stachel's conjecture

John Stachel advances the following conjecture (Stachel 1989a, 304-308; Stachel, 2007a, 261-292, 265; Stachel 2007b, 438-439).

It seems at least possible that the reason why Einstein gave up the November tensor $T_{il}^x$ in favor of the *Entwurf* field equations may lie in Einstein's belief that in the weak-field approximation space should be flat. Did the spatial metric of a static gravitational field cause Einstein to move down a slippery slope toward non-covariant field equations? In his paper, "Einstein's Search for General Covariance 1912 – 1915", Stachel argues that Einstein chose a pathway, which followed the mathematical physical knowledge of his day. Einstein expected that the Ricci tensor should reduce in the limit of weak-fields to his static gravitational field theory from 1912 and then to the Newtonian limit, if the static spatial metric is flat. This statement, says Stachel, appears to have led Einstein to reject the Ricci tensor, and "fall into the trap" of *Entwurf* limited generally covariant field equations.

Einstein, as conjectured by Stachel, had good reason for his long-held intuition that, in Newtonian theory space (as opposed to space-time) should be flat. The metric tensor in the special relativity limit is diag $(-1, -1, -1, c^2)$. The three space components are equal to –1 and the time component is equal to $c^2 = 1$. Within his static gravitational theory, Einstein replaced the gravitational potential by the variable speed of light. In his *Entwurf* theory he "formulated" an *Ansatz*: The same type of metric tensor degeneration appears in special relativity and in static gravitational fields, only that in the latter case $g_{44} = c^2$ is a function of the spatial coordinates. That is to say, the limit of weak static gravitational fields is diag $(-1, -1, -1, c^2(x, y, z))$. Therefore, three-dimensional space would remain Euclidean in these static fields. Einstein thus held a conception that, in the weak-field approximation, the spatial metric of a weak gravitational field must be flat. He obtained the Newtonian results by using the weak-field approximation and by assuming diag $(-1, -1, -1)$. Stachel concludes that Einstein decided to "keep his *Ansatz* for static metrics, and drop the Ricci tensor field equations".

We can yield important evidence for Stachel's conjecture. On page 20L of the *Zurich Notebook* Einstein modified the weak-field equations and added a term with the trace of the stress-energy tensor on the right-hand side of the equations [see equation (41) in Section 3.2]. He introduced the second term on the right-hand side of the weak-field equations (1) in such a way that he could now use the harmonic coordinate condition to satisfy both conservation of momentum-energy and the Newtonian limit. However, he finally crossed out these new weak-field equations because they did not allow the spatially flat metric that represented static gravitational fields (in Section 3.2 I further examine this example).

On page 21R Einstein wanted to check whether his spatially flat metric that represented static gravitational fields was compatible with Galileo's experimental law of free fall and the equivalence principle. He arrived at the fallacious conclusion that it was essential to Galileo's law of free fall. According to Galileo's experimental law of free fall, all bodies fall with the same acceleration in a given gravitational field. According to special relativity, the inertial mass is proportional to energy. If one mass fell differently from all others in the gravitational field then, with the help of this mass an observer in free fall (which for him locally his system is inertial) could discover that he was falling in a gravitational field. In this case, the acceleration of the falling mass would not be independent of the internal energy of the system and this violated special relativity (the inertia of energy). Therefore, Einstein concluded (i.e. "formulated" his *Ansatz*): According to the principle of equivalence, locally for the free falling observer and for an observer in a weak static field, the metric was represented by the form diagonal $(-1, -1, -1, c^2)$. However, the new weak-field equations from page 20L no longer provided a solution with a metric of this form and thus Einstein gave up his new field equations (Einstein 1912a, 442, 444, 447, 449; Janssen, Renn, Sauer, Norton and Stachel, 2007, 605-606, 633; 641).

Later in 1913, as indicated in the mathematical part of the *Entwurf* paper, Grossmann wrote: "The extraordinary importance of these conceptions [the Ricci tensor] for the power of *differential geometry* of a line element that is given by its manifold makes it a priori probable that these general differential tensors may also be of importance for the problem of the differential equations of a gravitational field… But it turns out that in the special case of the infinitely weak, static gravitational field this tensor does *not* reduce to the [Newton-Poisson equation]" (Einstein and Grossmann, 1913, 35-36). Grossmann appeared to have been influenced by Einstein's conception that, in the weak field approximation, the spatial metric of a static gravitational field must be flat.

Why did it take Einstein three years to arrive at the realization that the November tensor $T_{il}^x$ is not incompatible with Newton's law? According to Stachel, the answer to this question depends on Einstein's *Ansatz*.

First, in 1913 Einstein found non-generally covariant field equations (the Einstein-Grossmann *Entwurf* field equations) based on the criteria that the field equations should generalize the Poisson equation, be invariant at least under linear transformations and the conservation laws for the total gravitational and non-gravitational energy and momentum should follow from the gravitational field equations. He also gave a sketch of how Newton's law of gravitation follows from the linear approximation of the *Entwurf* field equations for the static case, a result he later demonstrated in detail in his 1913 lecture in Vienna (Einstein 1913, 1259). In the Vienna-lecture paper Einstein argued that in the weak-field case, the metric $g_{\mu\nu}$ adopts the following form:

$g_{\mu\nu} = \eta_{\mu\nu} + g_{\mu\nu}^*$.

The first term is the spatially flat metric that represents special relativity:

(20) $\eta_{\mu\nu} = \text{diag} (-1, -1, -1, c^2)$.

The second term $g_{\mu\nu}^*$ represents a small deviation of the metric (of the first-order approximation) from the constant values of $\eta_{\mu\nu}$.

The *Entwurf* field equations reduce to:

$$(21) \quad \Box g^*_{\mu\nu} = \frac{\partial^2 g^*_{\mu\nu}}{\partial x^2} + \frac{\partial^2 g^*_{\mu\nu}}{\partial y^2} + \frac{\partial^2 g^*_{\mu\nu}}{\partial z^2} - \frac{1}{c^2}\frac{\partial^2 g^*_{\mu\nu}}{\partial t^2} = \kappa T_{\mu\nu}.$$

Einstein made the following assumptions: He assumed a static gravitational field produced by a pressureless, static cloud of dust of mass density $\rho_0$ and he assumed (as a boundary condition) that the $g^*_{\mu\nu}$ vanish at infinity. In addition, all terms of $T_{\mu\nu}$ except for the $T_{44}$ term are neglected. $T_{44}$ is then identified with $\rho_0$, the mass density appearing in the Poisson equation. We then obtain from (21) Poisson's equation.

The spatial metric remains flat in this approximation:

(22) diag $(-1, -1, -1, c^2(x, y, z))$,

by solving:

$(22a)\ \Delta g^*_{44} = \kappa c^2 \rho_0,\ \text{for}\ \mu, \nu = 4,\ \text{and:}$

$(22b)\ \Delta g^*_{\mu\nu} = T_{\mu\nu} = 0,\ \text{for}\ \mu, \nu \neq 4,$

$\Delta$ is the Laplacian operator.

From (22a) and (22b) it follows that:

$(23)\ \ g^*_{\mu\nu} = 0,\ \text{for}\ \mu, \nu \neq 4,\ \text{and}$

$(23a)\ g^*_{44} = \frac{\kappa c^2}{4\pi} \int \frac{\rho_0 dv}{r} = 2G \int \frac{\rho_0 dv}{r},\ \text{for}\ \mu, \nu = 4,$

where (24) $\kappa = G\frac{4\pi}{c^2}$.

In his 1913 *Entwurf* paper, Einstein wrote that in special relativity, a material particle on which no forces are acting follows a straight line through flat space-time (Einstein and Grossmann 1913, 4-7):

$(25)\ \delta\left\{\int ds\right\} = 0,$

where the action is:

$(25a)\ s = \int L dt,$

and the Lagrangian is given by:

$(25b)\ L = -m\frac{ds}{dt},$

with $m$ the rest mass of the material point.

In the case of weak static gravitational fields, the equation of motion in the Newtonian limit is then obtained from equations (23a), (25), (25a), and (25b):

(26) $\dfrac{d^2 x}{dt^2} = -\dfrac{1}{2}\dfrac{\partial g_{44}^*}{\partial x}$.

Second, in June 1913, Einstein and Michele Besso calculated Mercury's perihelion and obtained a final disappointing result of 18" in the *Einstein-Besso manuscript*. They were looking for the equation of Mercury moving along the geodesic line in the static gravitational field of the Sun. In a very great distance from the Sun the gravitational field is so weak that it is not felt and we arrive back at the Minkowski flat metric. Einstein calculated the metric field of the Sun using the *Entwurf* vacuum field equations in a first-order approximation [see equation (22a)] and found that the static spatial metric was flat [see equation (22)]. Thus to first-order, Mercury moves along a geodesic curve that depends on flat components of the metric.

Janssen has written: "Einstein did not give up the Einstein-Grossmann theory once he had established that it could not fully explain the Mercury anomaly. He continued to work on the theory and never even mentioned the disappointing result of his work with Besso in print. So Einstein did not do what the influential philosopher Sir Karl Popper claimed all good scientists do: once they have found an empirical refutation of their theory, they abandon that theory and go back to the drawing board" ("The Einstein-Besso Manuscript: Looking Over Einstein's Shoulder", 12). Einstein could thus stubbornly stick to his *Entwurf* theory and to his *Ansatz*.

Two years later, in November 1915, Einstein transferred the basic framework of the calculation from the *Einstein-Besso manuscript* and corrected it according to his generally covariant November 1915 vacuum field equations. He tried to solve the November 1915 vacuum field equations to attain the perihelion advance of Mercury in the field of the static Sun. He calculated the metric field of the Sun using the November 1915 vacuum field equations in a first-order approximation. He discovered that the static spatial metric need not be flat and Mercury moves along a geodesic curve that depends on non-flat components of the metric.

Indeed, after presenting the final general theory of relativity, Einstein told Besso that (in 1912): "…Grossmann and I believed that the conservation laws were not satisfied and Newton's law did not result in first order approximation. You will be surprised by the occurrence of the $g_{11}...g_{33}$'s" (Einstein to Besso, December 10, 1915, *CPAE 8*, Doc. 162), that is to say, by the occurrence of non-flat components of the first order metric. This quote provides strong evidence in support of Stachel's conjecture. Einstein could not guarantee that the restricted November tensor $T_{il}^x$ be compatible with the law of energy-momentum conservation and reduce to the Newtonian limit. This problem is then tightly interwoven with Einstein's 1912-1914 belief that the static spatial metric is flat.

Stachel points out that later, however, it was claimed that one could not properly take the Newtonian limit of general relativity without the concept of an affine connection, and the corresponding affine reformulation of Newtonian theory.

The connection used in general relativity, called the Levi-Civita connection, is the Christoffel symbols of the second kind, equation (5). In 1923 Élie Joseph Cartan found a general affine connection which was non-flat, and it was still not the metric connection. The Christoffel symbols do not serve as the components of this connection and it is not written in terms of the metric tensor components.

Stachel explains that in the absence of the affine approach, more-or-less heuristic detours through the weak-field, special-relativistic limit (i.e. fast motion) followed by a slow motion approximation basically out of step with the special-relativistic approach, had to be used to obtain the desired Newtonian results.

Cartan defined the Ricci tensor in terms of the above non-flat connection, which could not break up into a Newtonian flat affine connection and a gravitational potential. In the local inertial system, which means in a freely falling system, we cannot separate gravity from inertia. This embodies the inertio-gravitational field. Accordingly, in the Newtonian limit, although we obtain Poisson's equation and space-time is flat, Cartan's affine connection remains non-flat, that is to say, the Ricci tensor is expressed in terms of a non-flat connection.

Stachel conjectures that had Einstein known about the affine connection representation of the inertio-gravitational field, he would have been able to see that the spatial metric can go to a flat Newtonian limit, while the Newtonian connection remains non-flat without violating the compatibility conditions between metric and connection.

Stachel argues that Einstein attempted to formulate in the best way he could his physical insights about gravitation and relativity already, and incorporate them in the equivalence principle. Einstein's attempt was hampered by the absence of the appropriate mathematical concepts. In 1907 Einstein still lacked the Riemanian geometry and the tensor calculus as developed by the turn of the century, i.e., based on the concept of the metric tensor; and later when he was using these, he then lacked more advanced mathematical tools (the affine connection); these could be later responsible for inhibiting him for another few years.

My comments on Stachel's conjecture: The theory put forward by Stachel has opened up new lines of investigation, particularly concerning this study. It also has the support of evidence from the *Zurich Notebook*. On page 22R, Einstein constructed a candidate generally covariant tensor, the November tensor $T_{il}^x$, from the Riemann curvature tensor. Having previously realized that the combination of the harmonic coordinate condition (2) and the Hertz restriction (2a) caused problems (because of the incompatibility with a spatial flat metric), on page 23L Einstein understood that the combination of the Hertz restriction (2a) and the $\vartheta$ restriction (12) allowed him to recover an expression from which the Newton-Poisson equation could be obtained as a first-order approximation (3c).

Nonetheless, the case is not yet solid. Einstein finally rejected the truncated November tensor $T_{il}^x$ because he could not reconcile the $\vartheta$ expression (12) with rotation. On the other hand, in 1915, he contracted the Ricci tensor in his November 4 field equations [the November tensor equal to the stress-energy tensor, multiplied by the gravitational constant, equation (4a)]. He then used the Hertz condition (2b) to rewrite his field equations (4a) in the form (3d), from which he obtained the Newton-Poisson equation as a first-order approximation. Thus, on November 4, 1915 Einstein first demonstrated that his field equations satisfy the conservation of energy-momentum (gravitational field + matter) and then, he used the Hertz condition (2b) (see Section 1). How did Einstein arrive at this point toward attaining a very successful and sophisticated ability? When he worked in 1912 he was unable to demonstrate that his field equations fulfill the conservation law and reduce to Poisson's equation. The key to the solution of his problem lay not only in Einstein's belief that in the weak-field approximation space should be flat, but also in additional factors: Towards 1915 Einstein became increasingly sophisticated – mathematically.

Indeed, Stachel argues that before 1915, Einstein attempted to formulate in the best way he could his physical insights about gravitation and relativity already, and incorporate them in the equivalence principle. However, his attempt was hampered by the absence of the appropriate mathematical concepts. One may ask, what are these mathematical concepts? Scholars try to answer this question. They say that these mathematical concepts of various kinds are, among others, coordinate conditions in the modern sense.

### 3.2. Jürgen Renn's and Michel Janssen's conjecture

Why did Einstein reject the November tensor $T_{il}^x$ in 1912-1913, only to come back to it in November 1915? Renn and Janssen form the following conjecture in connection with the above question (Janssen and Renn 2007, 849-850, 854-855, 858-861, 864, 880, 886, 900, 903-909; 2015, 31-33, 36).

Janssen and Renn mention Abraham Pais who writes on Einstein's 1913 work with Grossmann in his biography of Einstein, *Subtle is the Lord*: "Einstein still had to understand that this freedom expresses the fact that the choice of coordinates is a matter of convention without physical content. *That* he knew by 1915 […]. We now also understand Grossmann's difficulty with the Newtonian limit" (Pais 1982, 222-223). Pais has suggested that Einstein and Grossmann did not know about coordinate conditions. Pais, however, did not examine the *Zurich Notebook*. What is likely to puzzle us is that Einstein had already written the Hertz condition (2b) in the *Zurich Notebook*. Following Renn's and Janssen's conjecture, here it is obvious, "Pais was not entirely wrong". Although we are dealing with the same mathematical formula, Einstein, however, wrote the Hertz restriction (2a).

Most informative, moreover, is Norton's 1984 pioneering study of the *Zurich Notebook* (see Section 3.4). Based on Norton's study, Janssen and Renn note that the rotation metric does not satisfy the Hertz restriction (2a), and this complicates the problem. Thus in 1912, Einstein gave up the November tensor $T_{il}^x$ because the rotation metric does not satisfy the Hertz restriction (2a) and the $\vartheta$ expression (12). Why was it a problem for Einstein that the rotation metric does not satisfy the Hertz condition? Einstein truncated the November tensor $T_{il}^x$ by imposing the Hertz restriction (2a) and obtained candidate field equations. He studied the covariance of these field equations by studying the covariance of the Hertz restriction (2a). Evidently Einstein wanted the rotation metric to be a solution of the vacuum field equations, so that he could interpret the inertial forces in a rotating frame of reference as gravitational forces (i.e. the equivalence principle to be satisfied). It is clear that the tensor $T_{il}^x$ vanishes for the rotation metric (15). Having truncated $T_{il}^x$ by imposing the Hertz restriction (2a), the rotation metric no longer vanished for what was left of the tensor $T_{il}^x$.

Clearly that which is suggested by Pais and Norton ought to be taken into consideration. We come here to Janssen's and Renn's central suggestion: It was thus hard for Einstein to recognize that the generally covariant field equations are a simple and natural generalization of Newton's law because three problems had become entangled with one another in the *Zurich Notebook*: 1) The field equations extracted from the November tensor $T_{il}^x$ reduce to the Newtonian limit for weak static fields, 2) They allow Minkowski space-time in rotating coordinates and 3) They satisfy energy-momentum conservation.

The first two problems 1) and 2) are easily solved separately: Einstein first demonstrated in his first November 1915 paper that his November tensor [November 4 field equations, equations (4a)] satisfy the conservation of energy-momentum (gravitational field + matter), equation (10) (Einstein 1915a, 784). With the Hertz condition (2b) he subsequently rewrote his field equations (4a) in the form (3d), from which he obtained the Newton-Poisson equation as a first-order approximation. Finally, Einstein noted at the end of his 1915 paper that his new theory of general relativity allows transformations to a rotating coordinate system. Returning to the November tensor, therefore, solved the problem of the rotation metric (Einstein 1915a, 786).

However, in the *Zurich Notebook* the first two problems had become entangled with one another and with the problem of energy-momentum conservation. The entanglement was the result of Einstein's employing the same restriction [the Hertz restriction (2a) on the second part of the Ricci tensor and then the $\vartheta$ restriction (12)] to reduce the field equations to an expression that could lead him to Newton's law of gravitation (3b) and (3c), guarantee energy-momentum conservation and allow transformation to rotating coordinates. "Einstein used coordinate conditions – not just in the *Zurich notebook* but throughout the reign of the *Entwurf* theory – in a one-size-fits-all fashion: The same coordinate condition suits all problems". Instead of following each problem in turn and unraveling them, the problems were interwoven. Einstein was not then using modern coordinate conditions (2b) but coordinate restrictions (2a).

Coordinate conditions only have to do the first of these three things (recover the Poisson equation for weak static fields). The three problems can thus be disentangled by switching from coordinate restrictions to coordinate conditions. Einstein, Janssen and Renn believe, "made this switch when he saw that field equations based on the November tensor can be made compatible with energy-momentum conservation by imposing just one weak coordinate restriction" (restriction to unimodular transformations) and by recovering the Poisson equation with a coordinate condition (2b) in the modern sense [equations (11), (11a), (11b)]. Once he adopted modern coordinate conditions and was able to separate the two sets of conditions (coordinate conditions and coordinate restrictions), he concluded that the November tensor, equation (1a), rejected in the *Zurich Notebook* was acceptable after all. "Not only could Einstein now decouple the problem of energy-momentum conservation from the problem of recovering the Poisson equation, he could also decouple the latter from the problem of rotation. It is this disentanglement of various conditions and requirements that we tried to capture in the title of our paper: 'Untying the knot'."

According to the conjecture of Renn and Janssen, it took Einstein three years to arrive at the realization that the November tensor $T_{il}^x$ is not incompatible with Newton's law because the *Entwurf* theory had a crucial role as scaffolding for building an arch or a bridge between physics and mathematics. On its basis, Einstein first calculated the Mercury perihelion motion, and then worked out the preliminary mathematical structure of general relativity. The Mercury calculation eventually helped him to solve the problem of the Newtonian limit. Einstein and Grossman set up a mathematical variational formalism for the *Entwurf* field equations, from which he could eventually (in 1915) derive the conservation of energy and momentum. By November 1915, Einstein had the scaffold torn down only to be led back to the 1912 November tensor $T_{il}^x$.

We have now covered Renn's and Janssen's main theory. We can fully understand how their suggestion may solve the problem only by turning to Einstein's own calculations.

Let us look at how the *Entwurf* theory can build a bridge between mathematics and physics. In 1913, Einstein together with Michele Besso calculated the Mercury perihelion motion on the basis of the *Entwurf* theory.

Mercury's perihelion motion and the Newtonian limit are related to the geodesic equation. On page 41R of the *Zurich Notebook*, Einstein started from the $x$-component of the Newtonian equation of motion for a particle of mass $m$, constrained to move on the surface $f(x, y, z)$:

(27) $m\dfrac{d^2x}{dt^2} = \lambda\,\dfrac{\partial f}{\partial x}.$

The right-hand side of this equation is the normal force that constrains the particle to move along the surface $f = 0$, and $\lambda$ describes the magnitude of the force.

Instead of this equation Einstein wrote:

$(27a)$ $\dfrac{d^2 x}{ds^2} = \lambda' \dfrac{\partial f}{\partial x}$.

Hence, the trajectory of a particle is a geodesic line on the curved surface $f = 0$. $ds$ is the line element:

$(28)$ $ds^2 = \displaystyle\sum_{\mu\nu} g_{\mu\nu} dx_\mu dx_\nu$,

Therefore, a material point moves in a gravitational field on a geodesic line in a four-dimensional spacetime. The geodesic, a curve of shortest distance, is determined by the variational principle (Einstein 1912a, 355, 357; Janssen, Renn, Sauer, Norton and Stachel, 2007, 593).

In his 1914 review article on his *Entwurf* theory, "The Formal Foundation of the General Theory of Relativity", Einstein extended equation (27a). The motion of the particle corresponds to a geodesic line in a four-dimensional spacetime. He wrote the geodesic equation:

$(29)$ $\dfrac{d^2 x_\sigma}{ds^2} = \displaystyle\sum_{\mu\nu} \Gamma^{\sigma}_{\mu\nu} \dfrac{dx_\mu}{ds} \dfrac{dx_\nu}{ds}$,

in terms of the Christoffel symbols of the second kind (5) (Einstein 1914, 1046). The geodesic equation expresses the second covariant derivative with respect to *s*, which represents the proper time. The solutions of the geodesic equation are geodesic lines in spacetime.

In 1915 Einstein showed that Mercury moves on a geodesic line according to (29) under the influence of the gravitational field of the static Sun, which is determined by $\Gamma^{\sigma}_{44}$ (the components of the gravitational field of the static Sun). Clearly in 1915 Einstein considered the Christoffel symbols (5) as the components of the gravitational field. He calculated the equations of the geodesic lines in the gravitational field of the Sun and compared them with the Newtonian equations of the orbits of the planets in the solar system.

Einstein assumed that the motion of the planet takes place with a velocity less than the velocity of light. Thus $dx_1$, $dx_2$, $dx_3$ are smaller than $dx_4$. If on the right-hand side of equation (29) we consider only the term $\mu = 4$, and we take into consideration the components of the gravitational field to first-order approximation $\Gamma^{\sigma}_{44}$ , then (Einstein 1915b, 835-837):

$(27b)$ $\dfrac{d^2 x_\sigma}{ds^2} = \Gamma^{\sigma}_{44} = -\dfrac{\alpha}{2} \dfrac{x_\sigma}{r^3}$ , $\qquad \sigma = 1,2,3$,

$(27c)$ $\dfrac{d^2 x_4}{ds^2} = 0$, $\quad (\sigma = 4)$,

from which we obtain Newton's second law:

(30) $\dfrac{d^2 x_\sigma}{dt^2} = -\dfrac{\partial \Phi}{\partial x_4}$,     $\Phi = -\dfrac{\alpha}{2r}$,

where $\Phi$ is the scalar potential and $\alpha = 2GM$. This is equation (26) from Section 3.1.

The *Entwurf* theory also served as a scaffolding to overcome Einstein's earlier problem with the conservation laws of energy and momentum. In 1914, Einstein and Grossmann set up a variational formalism for the *Entwurf* theory from which it was easy to derive the conservation laws. That formalism was general enough to allow the derivation of conservation laws of energy and momentum also for the November tensor discarded by Einstein in the winter of 1912-1913. Having constructed such formalism, the variational formalism, is what allowed Einstein to switch from the *Entwurf* theory to the theory he presented on November 4, 1915. In the end the transition from the *Entwurf* theory to the November 1915 theory seems to have been a rather simple step; but what made this step possible was the scaffolding represented by the variational formalism Einstein had built for the *Entwurf* theory. He changed one element in the variational formalism developed for the *Entwurf* field equations: He redefined the gravitational field in terms of the Christoffel symbols (5).

Einstein wrote the 1914 Lagrangian L (Einstein 1914, 1076-1077; 1915a, 784):

(31) $\mathrm{L} = \sum\limits_{\mu\rho\tau\nu} g^{\tau\nu}\, \Gamma^{\rho}_{\mu\tau}\Gamma^{\mu}_{\rho\nu}$,

where the components of the gravitational field were equal to:

(31a) $\Gamma^{\tau}_{\nu\sigma} = \dfrac{1}{2}\sum\limits_{\mu} g^{\tau\mu}\dfrac{\partial g_{\mu\nu}}{\partial x_\sigma}$.

The generally covariant equations that represent the energy-momentum balance for matter in a gravitational field are given by:

(32) $\sum\limits_{\nu}\dfrac{\partial \mathfrak{T}^{\nu}_{\tau}}{\partial x_\nu} = \dfrac{1}{2}\sum\limits_{\mu\tau\nu} g^{\tau\mu}\dfrac{\partial g_{\mu\nu}}{\partial x_\sigma}\mathfrak{T}^{\nu}_{\tau} + \mathfrak{K}_{\sigma}$,

where this equation represents the divergence of the stress-energy tensor density $\mathfrak{T}^{\nu}_{\tau}$ and $\mathfrak{K}_{\sigma}$ represents outer forces acting on the system. Einstein considered $\mathfrak{K}_{\sigma} = 0$, thus:

(32a) $\sum\limits_{\nu}\dfrac{\partial \mathfrak{T}^{\nu}_{\tau}}{\partial x_\nu} - \dfrac{1}{2}\sum\limits_{\mu\tau\nu} g^{\tau\mu}\dfrac{\partial g_{\mu\nu}}{\partial x_\sigma}\mathfrak{T}^{\nu}_{\tau} = 0$.

The stress-energy tensor density $\mathfrak{T}_\tau^\nu$ is a symmetrical tensor of rank two, which is represented by a mixed stress-energy tensor density, i.e. $T_\tau^\nu \sqrt{-g} = \mathfrak{T}_\tau^\nu$. It relates the stress components with the mass sources and energy and momentum densities (Einstein 1914, 1056).

<u>Firstly</u>, in his first November paper of 1915, Einstein said that the importance of the action of the gravitational field on material processes led him to choose (31a) to represent the components of the gravitational field. Certainly then the trouble began and indeed in 1915 Einstein confessed that this was a "fateful prejudice" (Einstein 1915a, 782). On November 28, 1915 Einstein explained to Arnold Sommerfeld why he had grown dissatisfied with the earlier non-covariant *Entwurf* field equations (Einstein to Sommerfeld, November 28, 1915, *CPAE* 8, Doc. 153). Einstein told Sommerfeld that the "key to the solution" was the realization that not the gradient of the metric tensor (31a) but the Christoffel symbols of the second kind (5), are to be regarded as the expression for the components of the gravitational field $\Gamma_{\sigma\beta}^\alpha$. Hence, $\Gamma_{\sigma\beta}^\alpha$ in (31) should be given by equation (5).

<u>Secondly</u>, on closer inspection, there is an additional element that is implicitly contained in the above description. In 1914 Einstein published a second paper with Grossmann in which they wrote that the gravitational equations, "Are covariant with respect to all admissible transformations of the coordinate systems, i.e., with respect to all transformations between coordinate systems which satisfy the conditions $B_\sigma$ [...] = 0". These four coordinate conditions were to hold in all those systems, which were adapted coordinate systems, and in which the Einstein-Grossmann *Entwurf* field equations were valid. Einstein thus posed a coordinate restriction on the *Entwurf* field equations, and the coordinate systems satisfying this coordinate restriction $B_\sigma = 0$ were the adapted coordinate systems for the gravitational field. Coordinate transformations between two such adapted coordinate systems were "justified" transformations. Thus, Einstein and Grossmann restricted the coordinate systems by the four coordinate conditions $B_\sigma = 0$, and they showed that these conditions were direct consequence of the gravitational equations and the conservation laws but the covariance of the equations was nevertheless far-reaching in these coordinate systems (Einstein and Grossmann 1914, 217, 219, 224).

In Einstein's 1914 review article, "The Formal Foundation of the General Theory of Relativity", Einstein again restricted the coordinate systems to adapted coordinate systems, using the same four coordinate conditions $B_\sigma = 0$ (Einstein 1914, 1070, 1074-1077):

$$(11e)\ B_\sigma = \sum_{\alpha\sigma\nu} \frac{\partial^2}{\partial x_\sigma\, \partial x_\alpha} \left( g^{\mu\nu}\, \frac{\partial \mathrm{L}\sqrt{-g}}{\partial g_\sigma^{\mu\nu}} \right) = 0, \qquad \text{where } g_\sigma^{\mu\nu} \equiv \frac{\partial g^{\mu\nu}}{\partial x_\sigma}.$$

The ten *Entwurf* field equations determined the ten functions $g_{\mu\nu}$, but the $g_{\mu\nu}$ also satisfied the four equations $B_\sigma = 0$. *L* is the Lagrangian.

Consider the *Entwurf* field equations:

$$(11f) - \sum_{\alpha\tau} \frac{\partial}{\partial x_\alpha} \left( g^{\nu\tau} \frac{\partial L\sqrt{-g}}{\partial g^\sigma_\alpha} \right) = \kappa \mathfrak{T}^\nu_\sigma + \sum_{\tau\alpha} \left( -g^{\nu\tau} \frac{\partial L\sqrt{-g}}{\partial g^{\sigma\tau}} - g^{\nu\tau}_\alpha \frac{\partial L\sqrt{-g}}{\partial g^{\sigma\tau}_\alpha} \right).$$

According to equation (11e), for adapted coordinate systems the divergence of the term on the left-hand side of the *Entwurf* field equations vanishes and therefore:

$$(11g) \sum_\nu \frac{\partial}{\partial x_\alpha} \left\{ \mathfrak{T}^\nu_\sigma + \frac{1}{k} \sum_{\tau\alpha} \left( -g^{\nu\tau} \frac{\partial L\sqrt{-g}}{\partial g^{\sigma\tau}} - g^{\nu\tau}_\alpha \frac{\partial L\sqrt{-g}}{\partial g^{\sigma\tau}_\alpha} \right) \right\} = \sum_\nu \frac{\partial}{\partial x_\alpha} (\mathfrak{T}^\nu_\sigma + t^\nu_\sigma) = 0.$$

Einstein thus used the very same four coordinate restrictions $B_\sigma = 0$ to guarantee conservation of momentum and energy (10).

Einstein, however, demonstrated that *two coordinate restrictions* $B_\sigma = 0$ and $S^\nu_\sigma = 0$ and his *Entwurf* field equations guarantee the energy-momentum conservation law (10) for matter and the gravitational field.

$S^\nu_\sigma = 0$ is derived from an equation for the *Entwurf* gravitational tensor $\mathfrak{G}_{\mu\nu}$:

$$(32b) \sum_{\nu\tau} \frac{\partial g^{\tau\nu}}{\partial x_\nu} \mathfrak{G}_{\sigma\tau} + \frac{1}{2} \sum_{\mu\nu} \frac{\partial g^{\mu\nu}}{\partial x_\sigma} \mathfrak{G}_{\mu\nu} = 0.$$

Einstein considered the energy-momentum balance for matter in a gravitational field (32a) and obtained the above equations of similar form for $\mathfrak{G}_{\mu\nu}$ (Einstein 1914, 1076-1077).

According to equation (11g):

$$(8e) \ t^\tau_\alpha = \frac{1}{\kappa} \sum_{\alpha\tau} \left( -g^{\nu\tau} \frac{\partial L\sqrt{-g}}{\partial g^{\sigma\tau}} - g^{\nu\tau}_\alpha \frac{\partial L\sqrt{-g}}{\partial g^{\sigma\tau}_\alpha} \right).$$

This is the first term on the left-hand side of equation (32b). Einstein then rewrote the second term on the left-hand side of equation (32b) as:

$$(8f) \ t^\nu_\sigma = -2\kappa t^\alpha_\sigma = \delta^\alpha_\sigma L\sqrt{-g} - \sum_{\mu\nu} g^{\mu\nu}_\sigma \frac{\partial L\sqrt{-g}}{\partial g^{\mu\nu}_\alpha},$$

I will not provide a full mathematical derivation here, since it is not relevant to the present work.[3] He combined the two terms and obtained:

$$(8g) \ S^\nu_\sigma = t^\tau_\alpha - t^\nu_\sigma = 0 \ \rightarrow \ t^\nu_\sigma = t^\tau_\alpha.$$

---

<u>Thirdly</u>, in his first November paper of 1915, Einstein rewrote equation (32) with ordinary tensors only. He thus imposed the condition that the square root of the negative determinant of the metric tensor be equal to one $\sqrt{-g} = 1$ (a restriction to unimodular transformations) on equation (32), i.e. on the energy-momentum balance for matter $T_\tau^\nu$ in a gravitational field:

$$(32c) \sum_\nu \frac{\partial T_\sigma^\nu}{\partial x_\nu} = \frac{1}{2} \sum_{\mu\tau\nu} g^{\tau\mu} \frac{\partial g_{\mu\nu}}{\partial x_\sigma} T_\tau^\nu + K_\sigma,$$

with the four-vector representing outer forces $K_\sigma = 0$ one obtains equation (8a).

<u>Fourthly</u>, in 1914 Einstein wrote the action integral $J$ (Einstein 1914, 1069):

$$(33) J = \int L\sqrt{-g}\, d\tau,$$

and offered a method in terms of invariant integrals: If we change coordinates, the integral will be of the same form due to the $\sqrt{-g}\, d\tau$. Including this term in the integral gives an invariant expression valid for any arbitrary coordinate system. The integral (33) is thus invariant under an arbitrary change of coordinates (i.e. it has the same form in any arbitrary coordinate system).

However, in his first November paper of 1915, Einstein adopted $\sqrt{-g} = 1$ as a postulate, so that only substitutions of determinant 1 are permitted. Since $\sqrt{-g} = 1$, then the following action integral:

$$(33a) \int L\, d\tau,$$

is an invariant expression valid for any arbitrary coordinate system, under the restriction to unimodular transformations.

Einstein wrote the variation and obtained the field equations with the sources (the stress-energy tensor) on the right-hand side (Einstein 1915a, 784):

$$(34a) \sum_\alpha \frac{\partial}{\partial x_\alpha} \left( \frac{\partial L}{\partial g_\alpha^{\mu\beta}} \right) - \frac{\partial L}{\partial g^{\mu\nu}} = -\kappa T_{\mu\nu}.$$

He multiplied these equations by $g_\sigma^{\mu\nu}$ with summation over the indices $\mu$ and $\nu$, and obtained:

$$(34b) \sum_{\alpha\mu\nu} \frac{\partial}{\partial x_\alpha} \left( g_\sigma^{\mu\nu} \frac{\partial L}{\partial g_\alpha^{\mu\nu}} \right) - \frac{\partial L}{\partial x_\sigma} = -\kappa T_{\mu\nu} g_\sigma^{\mu\nu}.$$

Equation (32c) is obtained from the definition of the covariant divergence of an arbitrary symmetric tensor. It leads to:

$$(8a') \quad \sum_\alpha \frac{\partial T_\sigma^\alpha}{\partial x_\alpha} = -\sum_{\alpha\beta} \Gamma_{\sigma\beta}^\alpha T_\alpha^\beta = \frac{1}{2} \sum_{\mu\tau\alpha} g^{\tau\mu} \frac{\partial g_{\mu\nu}}{\partial x_\sigma} T_\tau^\alpha = -\frac{1}{2} \sum_{\mu\nu} \frac{\partial g^{\mu\nu}}{\partial x_\sigma} T_{\mu\nu}.$$

Equations (8a'), (34b), and the stress-energy components of the gravitational field $t_\sigma^\alpha$:

$$(8h) \quad -2\kappa t_\sigma^\alpha = \delta_\sigma^\alpha L - \sum_{\mu\nu} g_\sigma^{\mu\nu} \frac{\partial L}{\partial g_\alpha^{\mu\nu}},$$

lead to conservation of energy-momentum (10). Equation (8h) is equation (8f) with $\sqrt{-g} = 1$.

Einstein told his close friend Paul Ehrenfest that the conservation of energy-momentum (10) was covariant under unimodular transformations $\sqrt{-g} = 1$ because both the main equations from which it was derived, equations (8a') and (8h) [equation (8)] were also invariant under unimodular transformations (Einstein to Ehrenfest, January 17 1916, *CPAE* 8, Doc. 182).

According to Renn and Janssen, in the end the transition from the *Entwurf* theory, based on the physical strategy, to the November 1915 theory, based on the sophisticated mathematics of the Riemann tensor, seems to have been a rather simple step. If $\Gamma_{\sigma\beta}^\alpha$ are given by equation (5) and the formalism is restricted to unimodular transformations (adopting $\sqrt{-g} = 1$ as a postulate), the field equations are exactly the ones based on the November tensor.

Let us return to Einstein. Finally, in his 1916 review paper, "The Foundation of the General Theory of Relativity", Einstein wrote his field equations (4c) in unimodular coordinates:

$$(4d) \quad \frac{\partial \Gamma_{\mu\nu}^\alpha}{\partial x_\alpha} + \Gamma_{\mu\beta}^\alpha \Gamma_{\nu\alpha}^\beta = -\kappa \left( T_{\mu\nu} - \frac{1}{2} g_{\mu\nu} T \right), \quad \sqrt{-g} = 1.$$

Contracting (4d) and using equation (8), Einstein arrived at:

$$(4e) \quad \frac{\partial}{\partial x_\alpha} \left( g^{\sigma\beta} \Gamma_{\mu\beta}^\alpha \right) = -\kappa \left[ \left( T_\mu^\sigma + t_\mu^\sigma \right) - \frac{1}{2} \delta_\mu^\sigma (T + t) \right].$$

Similarly, from (4d) Einstein obtained:

$$(4f) \quad \frac{\partial}{\partial x_\alpha} \left[ \left( g^{\sigma\beta} \Gamma_{\mu\beta}^\alpha \right) - \frac{1}{2} \delta_\mu^\sigma g^{\lambda\beta} \Gamma_{\lambda\beta}^\alpha \right] = -\kappa \left( T_\mu^\sigma + t_\mu^\sigma \right).$$

Equations (4e) guarantee conservation of energy-momentum if the divergence of the left-hand side of (4f) is set equal to zero:

$$(4g) \quad \frac{\partial^2}{\partial x_\alpha \, \partial x_\sigma} \left[ \left( g^{\sigma\beta} \Gamma_{\mu\beta}^\alpha \right) - \frac{1}{2} \delta_\mu^\sigma g^{\lambda\beta} \Gamma_{\lambda\beta}^\alpha \right] \equiv 0.$$

The above equation gives the contracted Bianchi identities:

(35) $R_{\mu\nu} - \dfrac{1}{2} g_{\mu\nu} \mathrm{R} = 0$,

valid for a coordinate system $\sqrt{-\mathrm{g}} = 1$, i.e. in unimodular coordinates, from which (10) can be obtained.

In his short paper, "Hamilton's Principle and the General Theory of Relativity", Einstein treated the variational principle and formulated the field equations without assuming $\sqrt{-\mathrm{g}} = 1$. Thus, conservation of energy-momentum, equation (10), is guaranteed if the divergence of the left-hand side of the field equations [equation (37a) below] is set equal to zero. The two *identities*, $S^{\nu}_{\sigma} = 0$ and the Bianchi identity, which we may write succinctly as $B_{\sigma} = 0$, guarantee energy-momentum conservation (Weinstein 2017, 453-459).

In his November 20, 1915, paper (published later in March 1916) "The Foundations of Physics", David Hilbert was the first to obtain Einstein's November 25 field equations in an axiomatic framework based on the extensive use of a metric variational principle using a Lagrangian density $\mathfrak{L}$. Hilbert actually formulated the field equations without assuming $\sqrt{-g} = 1$. On November 20, 1915, Hilbert was the first to write the Ricci scalar (the curvature scalar) as the Lagrangian density (Hilbert 1915, 30, 38, 40).

In his short paper, "Hamilton's Principle and the General Theory of Relativity", Einstein made as few restrictive assumptions as possible. Following Hilbert, Einstein rewrote his *Entwurf* action integral (33) as follows (Einstein 1916b, 1112-1116):

(33b) $\displaystyle\int \mathfrak{L}\, d\tau$,     where  $L = \dfrac{\mathfrak{L}}{\sqrt{-g}}$,

and divided the Lagrangian density $\mathfrak{L}$ into two parts, one belonging to the gravitational field and the other to matter: $\mathfrak{L} = \mathfrak{G} + \mathfrak{M}$. $\mathfrak{G}$ depends on the components of the metric tensor $\mathrm{g}^{\mu\nu}$ and on their derivatives $\mathrm{g}^{\mu\nu}_{\sigma}$, and $\mathfrak{M}$ depends on the components of the metric tensor, and on the components of stress-energy tensor and its derivatives. However, $\mathfrak{L}$ also depends on $\mathrm{g}^{\mu\nu}_{\sigma\tau} = \dfrac{\partial^2 \mathrm{g}^{\mu\nu}}{\partial x_{\sigma}\, \partial x_{\tau}}$.

The functions:

(33c) $L = \dfrac{\mathfrak{L}}{\sqrt{-g}}, G = \dfrac{\mathfrak{G}}{\sqrt{-g}}$ and $M = \dfrac{\mathfrak{M}}{\sqrt{-g}}$,

are invariants under arbitrary transformations of the space-time coordinates.

Hence, (33b) is also an invariant. The variation of this action according to the variational principle:

$$\int \mathfrak{L}\, d\tau = 0,$$

leads to generally covariant field equations. Einstein then derived the field equations satisfied by $\mathfrak{L}$. He wrote the action:

$$\int \mathfrak{L}\, d\tau = \int \mathfrak{L}^*\, d\tau + F.$$

$\mathfrak{L}^*$ does not depend on $g_{\sigma\tau}^{\mu\nu}$. Thus the integral $\int \mathfrak{L}^*\, d\tau$ is not an invariant. Einstein added $F$ an additional integral that depends on the components of the metric tensor and on their derivatives.

He wrote the action in the following form (he took into account only the gravitational part of the Lagrangian):

$$\int \mathfrak{G}\, d\tau = \int \mathfrak{G}^*\, d\tau + F.$$

Today, the above equation is called the "Einstein-Hilbert action". Einstein, however, discovered that in the above expression the integral $\int \mathfrak{G}^*\, d\tau$ was not an invariant.

In his November 20, 1915, paper Hilbert illustrated that the Riemann curvature tensor $B_{\mu\nu\tau}^{\tau}$ allows the following invariant:

$$K = g^{\mu\nu} B_{\mu\nu\tau}^{\tau} = \frac{\mathfrak{G}}{\sqrt{-g}}$$

$K$ is the Ricci scalar (i.e. the scalar of the Riemann curvature tensor).

Following Hilbert, Einstein wrote an expression for $\mathfrak{G}$ in terms of the Ricci scalar and wrote the variation of the Einstein-Hilbert action (without $F$) in the following way:

$$(36)\ \delta \left\{ \int \frac{\mathfrak{G}}{\sqrt{-g}} \sqrt{-g}\, d\tau \right\} = \delta \left\{ \int \mathfrak{G}^*\, d\tau \right\}.$$

Since $\frac{\mathfrak{G}}{\sqrt{-g}}$ is an invariant, the left-hand side of the equation is invariant, and the right-hand side is invariant as well. $\int \mathfrak{G}^*\, d\tau$ is now invariant. Einstein then wrote the action and the variational principle, i.e. the right-hand side of equation (36) as:

$$\delta \left\{ \int \mathfrak{G}^*\, d\tau \right\} = 0.$$

We vary with respect to the metric tensor:

$$\int \delta g^{\mu\nu} \left\{ \frac{\partial}{\partial x_{\sigma}} \left( \frac{\partial \mathfrak{G}^*}{\partial g_{\sigma}^{\mu\nu}} \right) - \frac{\partial \mathfrak{G}^*}{\partial g^{\mu\nu}} \right\} d\tau.$$

This expression is also an invariant.

We obtain the field equations as the Euler-Lagrange equations:

$$(37) \quad \frac{\partial}{\partial x_\sigma}\left(\frac{\partial \mathfrak{G}^*}{\partial g_\sigma^{\mu\nu}}\right) - \frac{\partial \mathfrak{G}^*}{\partial g^{\mu\nu}} = \frac{\partial \mathfrak{M}}{\partial g^{\mu\nu}}.$$

We multiply (37) by $g^{\mu\nu}$ (contract) and obtain:

$$(37a) \quad \frac{\partial}{\partial x_\alpha}\left(g^{\mu\nu}\frac{\partial \mathfrak{G}^*}{\partial g_\alpha^{\mu\sigma}}\right) = -(\mathfrak{T}_\sigma^\nu + t_\sigma^\nu),$$

where $\mathfrak{T}_\sigma^\nu = -\frac{\partial \mathfrak{M}}{\partial g^{\mu\nu}}g^{\mu\nu}$,

and the stress-energy pseudo-tensor of the gravitational field is:

$$(38) \quad t_\sigma^\nu = \frac{1}{2}\left(\delta_\sigma^\nu \mathfrak{G}^* - g_\sigma^{\mu\alpha}\frac{\partial \mathfrak{G}^*}{\partial g_\nu^{\mu\alpha}}\right),$$

$$(38a) \quad t_\sigma^\nu = -\left(g^{\mu\nu}\frac{\partial \mathfrak{G}^*}{\partial g^{\mu\sigma}} + g_\alpha^{\mu\nu}\frac{\partial \mathfrak{G}^*}{\partial g_\alpha^{\mu\sigma}}\right).$$

Recall that the generally covariant equations that represent the energy-momentum balance for matter in a gravitational field are (32b):

$$(39) \quad \frac{\partial g^{\mu\nu}\mathfrak{G}^*}{\partial x_\nu} + \frac{1}{2}\frac{\partial g^{\mu\nu}\mathfrak{G}^*}{\partial x_\sigma} = 0.$$

Using equations (38) and (38a), equation (39) is brought into the form of (8g):

$$(40) \quad S_\sigma^\nu = t_\alpha^\tau - t_\sigma^\nu = 0.$$

*However, according to the above derivation equations (38) and (38a) come out of the blue!* In Einstein's *Entwurf* theory, equations (38) and (38a) are obtained from the *Entwurf* field equations [see equations (8e) and (8f)]. In other words, we need to derive equations (38) and (38a) on the basis of the 1916 general theory of relativity. Thus, the conclusion is that we cannot yet write down equation (40)!

How did Einstein obtain equations (38) and (38a)? He took into consideration *the principle of relativity and general covariance*. He considered the change $\Delta x_\nu$ due to infinitesimal transformation at some point of spacetime:

$$x_\nu' = x_\nu + \Delta x_\nu,$$

where the $\Delta x_\nu$ vanish outside some arbitrarily chosen region of spacetime.

For the components of the metric tensor $g^{\mu\nu}$ and their derivatives $g^{\mu\nu}_\sigma$ he considered the following transformation laws:

$$\Delta g^{\mu\nu} = g^{\mu\alpha} \frac{\partial \Delta x_\nu}{\partial x_\alpha} + g^{\nu\alpha} \frac{\partial \Delta x_\mu}{\partial x_\alpha},$$

$$\Delta g^{\mu\nu}_\sigma = \frac{\partial \Delta g^{\mu\nu}}{\partial x_\sigma} - g^{\mu\nu}_\alpha \frac{\partial \Delta x_\alpha}{\partial x_\sigma}.$$

He calculated $\Delta\mathfrak{G}^*$. An added difficulty was a term of the form $g^{\mu\nu}_{\sigma\tau} = \frac{\partial^2 \Delta x_\sigma}{\partial x_\nu \, \partial x_\alpha}$. $\mathfrak{G}^*$ depends only on the components of the metric tensor $g^{\mu\nu}$ and on their derivatives $g^{\mu\nu}_\sigma$. Einstein found:

(36b) $\Delta\mathfrak{G}^* = S^\nu_\sigma \dfrac{\partial \Delta x_\sigma}{\partial x_\nu} + 2 \dfrac{\partial \mathfrak{G}^*}{\partial g^{\mu\sigma}_\alpha} g^{\mu\nu} \dfrac{\partial^2 \Delta x_\sigma}{\partial x_\nu \, \partial x_\alpha},$      with

$$S^\nu_\sigma = 2 \left( \frac{\partial \mathfrak{G}^*}{\partial g^{\mu\sigma}} g^{\mu\nu} + \frac{\partial \mathfrak{G}^*}{\partial g^{\mu\sigma}_\alpha} g^{\mu\sigma}_\alpha \right) + \delta^\nu_\sigma \mathfrak{G}^* - \frac{\partial \mathfrak{G}^*}{\partial g^{\mu\alpha}_\nu} g^{\mu\alpha}_\sigma.$$

From this we obtain equations (38) and (38a):

(38b) $t^\nu_\sigma = \dfrac{1}{2} \left( \delta^\nu_\sigma \mathfrak{G}^* - \dfrac{\partial \mathfrak{G}^*}{\partial g^{\mu\alpha}_\nu} g^{\mu\alpha}_\sigma \right) = - \left( \dfrac{\partial \mathfrak{G}^*}{\partial g^{\mu\sigma}} g^{\mu\nu} + \dfrac{\partial \mathfrak{G}^*}{\partial g^{\mu\sigma}_\alpha} g^{\mu\sigma}_\alpha \right).$

Equation (36) can be written as follows:

$$\Delta \int \mathfrak{G} \, d\tau = \Delta \int \mathfrak{G}^* \, d\tau = 0.$$

Einstein concluded that $\Delta\mathfrak{G}^* = 0$ and by (38b) $S^\nu_\sigma = 0$. Putting these values in equation (36b), we obtain:

$$\int \frac{\partial \mathfrak{G}^*}{\partial g^{\mu\sigma}_\alpha} g^{\mu\nu} \frac{\partial^2 \Delta x_\sigma}{\partial x_\nu \, \partial x_\alpha} d\tau = 0,$$

from which we obtain the Bianchi identity:

$$\frac{\partial^2}{\partial x_\nu \, \partial x_\alpha} \left( g^{\mu\nu} \frac{\partial \mathfrak{G}^*}{\partial g^{\mu\sigma}_\alpha} \right) = 0,$$

which has the same mathematical form as the 1914 *Entwurf* equation (11e) $B_\sigma = 0$.

Conservation of energy-momentum, equation (10), is guaranteed if the divergence of the left-hand side of equation (37a) is equal to zero. The two identities $S^\nu_\sigma = 0$ and $B_\sigma = 0$ guarantee energy-momentum conservation. Einstein thus finally corrected his *Entwurf* theory.

Renn and Janssen have suggested that in 1914 Einstein had shown that the two conditions, $S^\nu_\sigma = 0$ and $B_\sigma = 0$, in conjunction with the *Entwurf* field equations guarantee energy-momentum conservation; but these conditions also determine the covariance of the *Entwurf* field equations, i.e. the class of "justified transformations" between adapted coordinate systems. In his review article (Einstein 1916a), Einstein had not connected the conditions (4g) guaranteeing energy-momentum conservation $B_\sigma = 0$ in unimodular coordinates to the corresponding covariance of the field equations. Instead, he had shown by direct calculation that these conditions are identically satisfied as long as unimodular coordinates are used. Renn and Janssen conclude that the variational treatment in arbitrary coordinates in (Einstein 1916b) fills two gaps. Einstein explicitly shows that energy-momentum conservation holds in arbitrary and not just in unimodular coordinates. More importantly, they say, it establishes for the new theory what Einstein had already found for the old 1914 *Entwurf* one, namely that there is an intimate connection between covariance and conservation laws.

My comments on Renn's and Jannsen's conjecture: Renn and Janssen have suggested that the *Entwurf* theory provided Einstein with the scaffolding for solving the problem with the Newtonian limit and the energy-momentum conservation law. In 1915 Einstein removed this scaffolding and returned to the November tensor. In fact, Einstein himself gave the following piece of advice. On May 28, 1953 he wrote to Maurice Solovine: "Moral: Unless one sins against logic, one generally gets nowhere; or, one cannot build a house or construct a bridge without using a scaffold which is really not one of its basic parts" (Einstein to Solovine, May 28, 1953, Einstein and Solovine 1993, 147).

As to coordinate conditions and coordinate restrictions. We will substantiate Renn's and Jannsen's conjecture by giving evidence from pages 19L-20L of the *Zurich Notebook*. Let us turn now to a few examples of the way that coordinate restrictions were being used by Einstein on pages 19L-20L. On page 20L Einstein wrote the two conditions, the Hertz condition, and the condition on the trace of the weak-field metric. He then, however, crossed them out because the combination of these two conditions caused problems: The trace of the weak-field metric was zero (Einstein 1912a, 442, 444). In order to avoid this problem, Einstein modified the weak-field equations (1), and added a term on the right-hand side of these equations with the trace of the stress-energy tensor:

$$(41)\ \Box g_{ik} = \rho_0 \frac{dx_i}{d\tau}\frac{dx_\kappa}{d\tau} - \frac{1}{4}\rho_0 \frac{dx_\kappa}{d\tau}\frac{dx_\kappa}{d\tau}.$$

Using the Kronecker delta this equation can be written more compactly as:

(41a) $\Box g_{ik} = \kappa \left( T_{ik} - \delta_{ik} \frac{1}{4} T \right)$.

The second term on the right-hand side was introduced in such a way that it partially solved the problem. These modified weak-field equations, however, did not allow the spatially flat metric (22) that Einstein continued to use to represent static fields. The problem was the following: Einstein's flat spatial metric of a static gravitational field was incompatible with field equations containing a term with the trace of the stress-energy tensor of matter.

Einstein thus crossed out the weak-field equations (41), and again modified the weak-field equations (1) from pages 19L and 19R in such a way that, he imposed the harmonic coordinate condition (2) *alone*, which ensured both the elimination of unwanted second-order derivative terms from the Ricci tensor and the vanishing of the divergence of the stress-energy tensor. Performing these two steps seems to support Renn's and Janssen's conjecture: these two steps had become entangled with one another. The entanglement was the result of Einstein's employing the same restriction (the harmonic restriction) to eliminate unwanted second-order derivative terms from the Ricci tensor and guarantee energy-momentum conservation.

With this strategy Einstein wrote the trace of the metric: $\sum_\kappa g_{\kappa\kappa} = U$. Previously, when Einstein imposed the two conditions (the Hertz and the harmonic coordinate conditions), he was obliged to add a term with the trace of the stress-energy tensor on the right-hand side of the weak-field equations. Now, he was imposing only the harmonic coordinate condition (2), and so he added a term with the trace of the metric on the left-hand side of the weak-field equations:

(42) $\Delta \left( g_{11} - \frac{1}{2} U \right) = T_{11}$, $\Delta g_{12} = T_{12}, \dots \Delta g_{14} = T_{14}$.

Using the Kronecker delta this equation can be written more compactly as:

(42a) $\Box \left( g_{ik} - \frac{1}{2} \delta_{ik} U \right) = T_{ik}$.

ensuring the vanishing of the divergence of the stress-energy tensor by imposing:

$\frac{\partial}{\partial x_\kappa} \left( g_{ik} - \frac{1}{2} \delta_{ik} U \right) = 0$,

Einstein's modified weak-field equations had removed the need for the Hertz condition (2a).

Einstein, however, wrote the modified weak-field equations in an alternative form. Taking the trace on both sides of equation (42), he found:

$-2\Delta U = \sum T_{\kappa\kappa} = \sum T$.

Einstein wrote the left-hand side term without a minus sign in front and he subsequently partly corrected his error. Inserting the above relation into equation (42), we arrive at:

$$(43) \ \Delta g_{11} = T_{11} - \frac{1}{2} \sum T_{\kappa\kappa}, \ \ \Delta g_{12} = T_{12}, \dots \Delta g_{14} = T_{14}.$$

The correct equations can be written more compactly as:

$$(44) \ \Box g_{ij} = T_{ij} - \frac{1}{2} \delta_{ij} \left( \sum T_{\kappa\kappa} \right).$$

Curiously, the left-hand side of equation (42a) is the linearized version of the Einstein tensor:

$$(35a) \ G_{\mu\nu} = R_{\mu\nu} - \frac{1}{2} g_{\mu\nu} R,$$

and equation (44) is the linearized version of equation (4c). See also equation (41a) above. There is no indication in the *Zurich Notebook* that Einstein tried to find the exact equations corresponding to these weak-field equations and the previous ones (41).

However, Einstein's flat spatial metric of a static gravitational field (22) was incompatible with field equations (43) and (44) containing a term with the trace of the stress-energy tensor. In addition, Einstein could not guarantee energy-momentum conservation and he made a fresh start with the whole calculation on page 21L (Janssen, Renn, Sauer, Norton and Stachel, 2007, 605-606, 633-635).

### 3.3. Jürgen Renn's and Tilman Sauer's Conjecture

Another answer that might be given to the question of why Einstein rejected equations of much broader covariance in 1912-1913 is Jürgen Renn's and Tilman Sauer's conjecture that, accepting the correct mathematical expression in 1912 required abandoning a few heuristic principles that Einstein could not yet reconcile with his field equations of 1912 (Renn and Sauer 2007, 123-125, 129, 147-149, 162, 217-219, 225, 243-246, 269, 271, 273; Janssen, Renn, Sauer, Norton and Stachel, 2007, 494; Janssen and Renn 2015, 35-36).

Renn and Sauer conjecture that, in 1912 Einstein searched for a gravitational theory, and a gravitational field equation, that would satisfy some heuristic requirements. Einstein had to cope with new mathematical tools, and at the same time he was guided by few heuristic principles. They formulate the heuristic principles that played a role in Einstein's search and rejection of generally covariant field equations between 1912 and 1913: The generalized principle of relativity, the equivalence principle, the correspondence principle (Newtonian limit) and the principle of conservation of energy and momentum.

I begin with the principle of equivalence. In 1907 and in 1911 Einstein formulated the "principle of equivalence". In his 1907 paper, "On the Relativity Principle and the Conclusions Drawn from It", Einstein assumed the complete physical equivalence of a homogeneous gravitational field and a corresponding (uniform) acceleration of the reference system (Einstein 1907, 454). In his 1911 paper, "On the Influence of Gravitation on the Propagation of Light", Einstein formulated the more mature (heuristic) equivalence principle. He obtained by theoretical considerations of the processes which take place relatively to a uniformly accelerating reference system, information as to the course of processes in a homogeneous gravitational field (Einstein 1911, 899-900). In 1912 Einstein developed the static gravitational field theory. He told Michele Besso that while developing this theory, the most interesting discovery was the following: He found that the principle of equivalence of acceleration and gravitation holds only for infinitely small systems (Einstein to Besso March 26, 1912, *CPAE* 5, Doc. 377).

Renn and Sauer have pointed out that after 1912, Einstein implemented the equivalence principle by letting the metric field $g_{\mu\nu}$ represent both the gravitational field and the inertial structure of space-time. The metric field is a solution of the same field equations in the coordinate systems of two observers. This is only automatically true if the field equations are generally covariant. The equivalence principle became the principle core of Einstein's search for a generalization of the relativity principle for non-uniform motion.

Renn and Sauer now discuss Einstein's generalized principle of relativity. Einstein attempted to generalize the principle of relativity by requiring that the covariance group of his new theory of gravitation be larger than the group of Lorentz transformations of special relativity. In his understanding this requirement was optimally satisfied if the field equation of the new theory could be shown to possess the mathematical property of general covariance. Einstein was thus under the impression that the principle of relativity for uniform motion of special relativity could be generalized to arbitrary motion if the field equations possessed the mathematical property of general covariance. And if the principle of relativity is generalized then the equivalence principle is satisfied, because according to this principle, an arbitrary accelerated reference frame in Minkowski space-time can precisely be considered as being physically equivalent to an inertial reference frame if a gravitational field can be introduced which accounts for the inertial effects in the accelerated frame. Einstein then tried to construct field equations of the broadest possible covariance.

Our next principle is the principle of conservation of energy and momentum. The energy-momentum conservation principle played a crucial role in Einstein's static gravitational theory of 1912. Einstein started with the relation between mass and energy from special relativity. He extended it within his theory of static gravitational fields. However, scientists (Max Planck and Max Laue) were already extending special relativity to relativistic dynamics, and conservation of energy and momentum centered at that time upon a four-dimensional stress-energy-momentum tensor. Starting in 1912 Einstein embodied the mass and energy relation in the stress-energy-momentum tensor as the source of the gravitational field. Einstein also required that the gravitational field equation should be compatible with the generalized requirement of energy and momentum conservation.

Finally, we come to the correspondence principle. Einstein was searching in two directions: He was looking for an equation of motion for bodies in a gravitational field and for a field equation determining the gravitational field itself, the generalization of the Poisson equation. The Poisson equation of classical gravitation theory describes how gravitating matter generates a gravitational potential. This potential can then be related to the gravitational field and to the force acting on the material particles exposed to it. Einstein's theory of static gravitational fields provided a starting point, since it represents a step beyond the Poisson equation towards a generalized relativistic theory.

On page 21R of the *Zurich Notebook* Einstein wanted to check whether his new gravitational field equations (1) fulfilled the requirement of recovering the familiar Newtonian gravitation theory (i.e. the Poisson equation of Newtonian theory for the scalar Newtonian potential) in the special case of the low velocity limit and weak static gravitational field. He expected that under appropriate limiting conditions, the theory he was developing would first reproduce the results of his 1912 theory of static gravitational fields; that is, he expected his new theory would reduce to his own earlier non-linear static gravitational field equation from March 1912, in which gravitational potential is represented by a variable speed of light (Einstein 1912b, 456):

(45) $c\Delta c - \frac{1}{2}(\text{grad}\,c)^2 = \kappa c^2 \sigma,$

where the second term on the left hand side of the equation is the energy density of the gravitational field multiplied by $c$ the velocity of light (acting as its own source), $\sigma$ denotes the mass density and the energy density, and $\frac{\kappa c^2}{2}$ denotes the gravitational constant $G$ [equation (24)].

Subsequently, under further constraints, Einstein would be able to recover equation (22a), the classical Poisson equation.

Renn and Sauer explain that from a qualitative point of view, Einstein constantly oscillated between what they have called "two strategies", that is to say, "the physical strategy" (how to cause the Newtonian limit to appear and guarantee that the field equations be compatible with energy-momentum conservation) and "the mathematical strategy" (constructing a generally covariant candidate for the left-hand side of the field equations). They maintain that each of Einstein's above heuristic principles against which constructions of candidates for field equations would have to be checked could be used either as a construction principle or as a criterion for their validity. In the *Zurich Notebook* Einstein started from mathematics (i.e. the mathematical strategy). He first tackled relativity and equivalence. He then arrived at general covariance (he constructed a candidate generally covariant tensor from the Riemann curvature tensor on page 22R), and then moved on to correspondence (Newtonian limit) and conservation of momentum and energy. Afterwards it was just the other way round (from pages 24R onwards). He gave up mathematics (i.e. the mathematical strategy) and established field equations – that would lead him to the so-called *Entwurf* field equations – while starting from physics (i.e. the physical strategy). He first tackled correspondence and conservation and then relativity and equivalence, and lost general covariance.

The interplay of the four heuristic principles with the new tools of absolute differential calculus of 1912 that Einstein was exploring governed the form of the field equations he was finally left with at the end of the *Zurich Notebook*. Consequently, at the end of the day Einstein's conditions overdetermined his research between 1912 and 1913.

Renn and Sauer have suggested that Einstein stubbornly expected that the theory he was developing would satisfy the correspondence principle. They provide the following example. The November tensor $T_{il}^x$ has a surprisingly elegant form. What is it made of? The divergence of the Christoffel symbols (5) plus a quadratic expression in the Christoffel symbols. Ironically, if the Christoffel symbols are taken to represent the gravitational field, the November tensor $T_{il}^x$ [or Einstein's candidate for the left-hand side of the field equations, (4a)] will have a canonical form. The left-hand side of equation (45) is exactly of this form:

(45a) $\dfrac{\Delta c}{c} - \dfrac{1}{2c^2}\text{grad}^2 c = \kappa\sigma,$

(the divergence of the gravitational field plus the gravitational field-squared).

Of course, Einstein could not have recognized that because such an interpretation was in conflict with his heuristics, which demanded the implementation of the correspondence principle first by imposing an appropriate coordinate restriction, the Hertz or/and the $\vartheta$ restriction. In this early stage, however, Einstein first imposed a coordinate restriction to meet the demands of the correspondence principle and then looked for an interpretation of a candidate in terms of gravitational field components. This is a tragicomedy of static field theory and loss of November tensor (in terms of the Christoffel symbols).

While searching for different candidates for the gravitational field equations, Einstein had an additional problem, to ensure the compatibility of the different heuristic requirements by integrating them into a coherent gravitation theory represented by a consistent tensorial framework. This complexity led to the following situation: The correspondence principle became the weightiest of Einstein's heuristic principles. At this stage Einstein explored a different candidate for the gravitational field equations, and implemented differently the correspondence principle. And in fact he eventually perused all options. Of course he could not give up so easily the conservation principle and the settings of the stress-energy tensor. He had to solve the conflict between them in order to solve the incompatibility problem between the correspondence and conservation principles. Einstein changed the left hand side of the field equations and solved the conflict between the two above principles, but had to check whether the conservation principle was fully satisfied. Once this issue was settled the correspondence principle caused problems with the theory of static gravitational fields (with Einstein's conception of the metric of static gravitational fields which has to be spatially flat).

At the end the match between the correspondence principle and the conservation principle was achieved at the expense of the generalized principle of relativity. At some stage, thus, Einstein appeared to somewhat forgot a little from the generalized principle of relativity; the starting point of his research project. Meanwhile he had developed many tools that would finally lead him to the goal of generally covariant field equations. However, at this stage he was not quite sure about the most important principles his novel theory should fulfill. He found it difficult to establish a match between the equations and the principles, between mathematics and physics.

During 1912-1913, Einstein gradually created the tensorial framework for his future general relativity. It appears that Einstein's long vacillating between general covariance and the correspondence principle is a symptom of his fixation on his older 1912 static theory of gravitational fields and his long-held belief that the static spatial metric was flat. Einstein needed an extra two years to gradually and mentally switch into a new paradigm; yet he would always envision Riemann's calculus in terms of his heuristic principles.

Between 1913 and 1914 Einstein adopted a physical strategy adapted to match the requirements of energy-momentum conservation and the generalized principle of relativity. It was thus an extension of the mathematical strategy. The insights Einstein had acquired pursuing the mathematical strategy in the *Zurich Notebook* set standards for his elaboration of the 1913 *Entwurf* theory developed along the lines of the physical strategy: in 1914 he derived the *Entwurf* field equations along the mathematical strategy. The basis for the derivation of the *Entwurf* field equations was provided by the variational formalism. In October 1915 Einstein realized that he had derived the non-covariant *Entwurf* field equations along the mathematical strategy, which was actually naturally suitable for generally-covariant field equations. After abandoning the *Entwurf* theory, Einstein returned to the November tensor he had derived along the mathematical strategy three years earlier in the *Zurich Notebook*. Renn and Sauer conclude that it was a return to the mathematical strategy applied to the absolute differential calculus (tensor calculus).

Renn and Janssen have added that in November 4, 1915 Einstein used the *Entwurf* theory as a scaffold to build the arch of the November 4, field equations. The *Entwurf* theory provided Einstein with the scaffolding for solving the problems with the Newtonian limit and conservation of momentum and energy. In the November 4, 1915 paper coordinate restrictions were turned into coordinate conditions and the relation between covariance and energy–momentum conservation was inverted from conservation restricting covariance to covariance guaranteeing conservation. Janssen and Renn argue that this was a success of the physical strategy.

It seems that in November 1915 both strategies, the physical and the mathematical, came into play. In October 1915, Einstein realized that his *Entwurf* Lagrangian function (31) did not uniquely lead to the *Entwurf* field equations. It must have been tempting for Einstein to look for other Lagrangians. However, in October-November 1915, say Renn and Sauer, Einstein took the mathematical strategy more seriously again when the following mathematical knowledge forced him to reconsider his physical prejudices: The components of the Christoffel symbols, equation (5), have a central role in the geodesic equation (29) and in the definition of the Riemann and Ricci tensors. It was not farfetched for Einstein to reinterpret the Christoffel symbols as representing the components of the gravitational field, a choice that almost immediately leads to the Lagrangian of the first November paper (31) (and of course to the understanding why the *Entwurf* Lagrangian function does not uniquely lead to the *Entwurf* field equations).

Einstein's earlier experience, documented in the *Zurich Notebook*, might have helped him to find this path from the *Entwurf* theory to the first November paper. In the *Zurich Notebook*, Einstein had already explored the relation between tensors (the Ricci tensor $T_{il}$) expressed in terms of the Christoffel symbols (and the November tensor $T_{il}^x$) and those expressed in terms of the derivatives of the metric, in the context of studying the $\vartheta$ expression, equation (12).

Until October-November 1915, Einstein considered the derivatives of the metric (gradient of the metric tensor), rather than the Christoffel symbols, as the true representation of the gravitational field. In October-November 1915, Einstein realized that not the first but the latter are to be regarded as the expression for the gravitational field.

My comment on Renn's and Sauer's conjecture: Einstein was caught in a vicious trap and was constantly oscillating between two strategies, the physical strategy and the mathematical strategy, and among four heuristic principles (relativity, equivalence, correspondence, and conservation of momentum and energy). Renn and Sauer add to and expand on the studies presented in Sections 3.1 and 3.2. They emphasize the qualitative aspect of Einstein's inner struggles. Renn's and Sauer's conjecture together with the two previous conjectures, that of Stachel and Renn and Janssen, provide a reasonably adequate account of Einstein's rejection of the November tensor $T_{il}^x$ in 1912-1913 and his return to it in November 1915.

## 3.4. John Norton's Conjecture

It seems that Renn's, Janssen's and Sauer's theories offer a promising path for interpretation. Here we must be extremely careful. Although this seems a likely solution, it should be pointed out that it is by no means obvious that Einstein's 1912 understanding of coordinate conditions was different from his 1915 understanding of these. There is other evidence, however, says John Norton, which is based on the same source, the *Zurich Notebook* that while it is certain that Einstein used coordinate restrictions to restrict the November tensor $T_{il}^x$, prior to it Einstein used coordinate conditions. No consensus has yet been reached as to Einstein's 1912 understanding of coordinate conditions.

Norton offers the following conjecture (Norton 1984, 255, 261, 265-266, 270, 274-276, 287-289; Norton 2000, 137-138, 147, 151, 155; Norton 2007, 721, 732, 734-735, 739, 741, 745-750, 758-763; Räz 2016, 180, 183, 189; Janssen, Renn, Sauer, Norton and Stachel, 2007, 605).

First, Norton agrees with Renn and Sauer (and Janssen) that as Einstein formulated and tested candidate field equations for his theory in the *Zurich Notebook*, he *consciously* alternated between two explicit strategies: the physical and the mathematical approaches. Einstein expected both to yield the same result so that one could be used to test the product of the other. When he erroneously concluded they did not yield the same results, he had to choose. He chose in favor of the first, the physical approach. Einstein had erroneously judged the November tensor, which actually satisfies all his conditions, as inadmissible. He would finally adopt this tensor at the end of 1915 but meanwhile in 1913 he chose the *Entwurf* theory and labored on, trying to convince himself of the admissibility of that theory.

Norton, however, suggests that towards the end of 1915 Einstein reversed his decision of 1913 and discovered that his dilemma was of his own making: "When [in the winter of 1912-1913] the physical requirements appeared to contradict the formal mathematical requirements, he had then chosen in favour of the former. He now [in 1915] chose the latter and, writing down the mathematically natural equations, found himself rapidly propelled towards a theory that satisfied all the requirements and fulfilled his 'wildest dreams'". Now there is a punch line. Through November 1915, Einstein was in intense competition with the mathematician David Hilbert. He *consciously* chose to allow the natural mathematical constructions of tensor calculus to guide him and rapidly bring him to the final result, and so the final field equations were communicated by an exhausted Einstein at the end of the month. Norton speaks about Einstein's "last minute reversal over the heuristic power of mathematics". He argues that "Einstein, even at this last moment, was quite disparaging and even scornful of the heuristic power of mathematical simplicity".

Norton argues that Einstein was indifferent and at times even derisive of considerations of mathematical simplicity early in his career. Pais reported that many years later Einstein told Valentine Bargmann (his assistant in Princeton) that before 1912 he regarded the transcription of his theory into tensor form as "überflüssige Gelehrsamkeit" (superfluous learnedness) (Pais 1982, 152). Einstein's disparaging attitude towards mathematics, says Norton, was only weakened temporarily in the winter of 1912-1913 by the need to proceed within a context more mathematically sophisticated than any in which he had worked before. However, Einstein's indifference to mathematical simplicity persisted up to and through the years in which he worked on completing the general theory of relativity.

I would like to add that the competition between Einstein and Hilbert culminated in November of 1915 when Einstein adopted a mathematical approach and Hilbert adopted a physical approach: In an ironic twist of fate, in the November 20 proofs of Hilbert's paper (which did not contain a generally covariant theory) Hilbert based his assertion on a slightly more sophisticated version of Einstein's hole argument against general covariance (after Einstein had silently dropped it), which he would eventually drop later when he would publish his paper in March 1916 (Stachel 1999, 358).

Second, regarding coordinate condition, already in 1984, John Norton had studied the *Zurich Notebook* and arrived at the conclusion that Einstein was certainly aware of the modern use of coordinate conditions in 1912; he was aware of the freedom to apply coordinate conditions to generally covariant field equations. Moreover, he even knew of the harmonic and Hertz coordinate conditions that could be used to reduce the Ricci tensor to a Newtonian form. What then led Einstein to reject the Ricci tensor?

Norton has pointed out that we know that in the case of Einstein's *Zurich Notebook* and *Entwurf* theory, Einstein was blinded by a misconception about static gravitational fields: Einstein's most favored special case was that of the static gravitational field, equation (22). He believed that the static spatial metric was flat. It is clear that Einstein's and Grossmann's rejection of the Ricci tensor need not be explained in terms of a simple error; rather it may have resulted from deep-seated misconceptions, especially Einstein's understanding of static gravitational fields, which was inconsistent with the Ricci tensor. Norton has noted that Einstein used the term "prejudice" – a belief not properly grounded in evidence (Einstein to Sommerfeld, November 28, 1915, *CPAE* 8, Doc. 153).

According to Norton, Einstein's explorations were based on the principle of equivalence, which asserted that a transformation to uniform acceleration in a Minkowski spacetime yielded a homogenous gravitational field. If one transforms from equation (20) to a coordinate system in uniform acceleration, the metric reverts to a form Einstein associated with a homogeneous gravitational field, which has the form of equation (22). Norton explains that Einstein's mistake, in 1912 and 1913, was that he considered the spatial flatness of the above metric of a homogeneous gravitational field a property of all static fields.

As already stated in Section 3.1, in June 1913 Einstein and Besso calculated Mercury's perihelion. Einstein calculated the metric field of the Sun using the *Entwurf* vacuum field equations in a first-order approximation and found that the static spatial metric was flat. Thus to first-order, Mercury moves along a geodesic curve that depends on flat components of the metric. Two years later, in November 1915 Einstein calculated the metric field of the Sun using the November 1915 vacuum field equations in a first-order approximation. He discovered that the static spatial metric need not be flat and Mercury moves along a geodesic curve that depends on non-flat components of the metric. Einstein told Besso that he was surprised by the occurrence of non-flat components of the first order metric. Norton has suggested that Einstein's flat static spatial metric, equation (22), was incompatible with field equations containing a term with the trace of the stress-energy tensor of matter (see example at the end of Section 3.2). Einstein had added such a trace term to the field equations (4c) of November 1915 *after* he had discovered that Mercury moves along a geodesic curve that depends on non-flat components of the metric, i.e. only after he had understood that the static spatial metric need not be flat.

Norton says that Einstein was troubled with another problem. He had one final misconception or "prejudice" about the form of his field equations in rotating coordinate systems, Minkowski spacetime when viewed from uniform rotating coordinates. He expected the November tensor $T_{il}^x$ to hold in the case of the rotation metric. Einstein imposed the Hertz condition (2a) on the tensor $T_{il}^x$, and recovered an expression that could lead him to Newton's law of gravitation, equation (3b) (see Section 1). Norton comes to the conclusion that Einstein dealt with the rotation metric in a manner that suggests that he expected the tensor $T_{il}^x$ to retain the form (3b) in rotating coordinate systems after the application of the appropriate coordinate condition. The Hertz condition (2a) "does not hold in this case, nor does the harmonic condition, another argument for Einstein against the Ricci tensor".

Renn and Janssen agree. Moreover, they have even adopted this conjecture (see Section 3.2). They argue that "Einstein's argument is cogent only if the conditions are seen as a coordinate restriction rather than a coordinate condition. In fact, it was in an attempt to make sense of these remarks in (Norton 1984) that one of us (JR [Renn]) first hit upon the distinction between coordinate conditions and what we have come to call coordinate restrictions" (Renn and Janssen 2007, 886).

Indeed, in 2007 Norton together with Stachel, Renn, Janssen, and Sauer produced the first systematic attempt to study equation by equation Einstein's *Zurich Notebook*. Their examination of the notebook revealed that Einstein discussed at length in relation to the November tensor $T_{il}^x$ coordinate restrictions; but then there began scholarly debate concerning Einstein's use of coordinate conditions. Renn, Janssen, and Sauer have suggested that Einstein was not aware of the modern usage of coordinate conditions at the time of the *Zurich Notebook* and that he rather only relied on so-called coordinate restrictions, which severely limit the covariance of the field equations and made the theory unacceptable. There can be little doubt that from page 22R on at least, Einstein used coordinate restrictions. On the other hand, Norton has argued that Einstein appears to have had the modern notion of coordinate conditions all along because he used coordinate conditions in the modern sense until page 22R of the *Zurich Notebook*; he was thus able to use coordinate conditions to bring the field equations extracted from the November tensor $T_{il}^x$ into correspondence with the Newton-Poisson equation. Norton's conjecture is based on his 1984 conjecture.

In 1989 Stachel discussed Norton's 1984 suggestion: "A study of Einstein's published papers and private correspondence between 1912–1915 convinced me that the standard explanation for his failure to arrive at the correct gravitational field equations until the end of this period [Pais 1982, 222] – namely, his presumed lack of understanding of the meaning of freedom of coordinate transformations in a generally covariant theory and the ability to impose coordinate conditions that this freedom implied – could not be correct […]. On the basis of his study of a research notebook [the *Zurich Notebook*] of Einstein from the early part of this period, John Norton was able to prove that Einstein already was aware of the possibility of imposing coordinate conditions on a set of field equations, and indeed had used the harmonic coordinate conditions" (Stachel 1989b, 168). Norton's conjecture is tightly interwoven with Einstein's belief that the static spatial metric must be flat. It is not surprising, therefore, to find Stachel examining Norton's theory.

During his work on the *Zurich Notebook*, Norton has suggested that page 22R of the notebook marks a turning point in how Einstein employed coordinate conditions. He has pointed out that there is clear evidence that Einstein used coordinate restrictions after page 22R of the *Zurich Notebook*. Prior to it, Einstein used coordinate conditions. For instance, at the bottom of page 19L, on which Einstein introduced the harmonic coordinate condition equation (2), he wrote: "Valid for coordinates which satisfy the equation $\Delta\varphi = 0$ [Laplace's equation]". Einstein thus imposed the harmonic coordinate condition on the Ricci tensor and recovered Laplace's equation in what we call the "harmonic coordinates". According to Norton, this provides support of the conjecture that, on page 19L Einstein used coordinate conditions. A coordinate restriction limits the covariance of the field equations. If the harmonic condition were used as a coordinate restriction, we might expect Einstein at some point to check its covariance in the way that he checked the covariance of the Hertz restriction (2a) and the $\vartheta$ expression (12). The pages 19L–21R of the *Zurich Notebook* contain no such check. On the other hand, one cannot assume that the $\vartheta$ expression is simply a coordinate condition being used to reduce the November tensor $T_{il}^{x}$ to the Newtonian form for the case of the Newtonian limit. First, if the $\vartheta$ expression is a coordinate condition, then there is no need to investigate its covariance under rotation transformations [equations (12), (17), and (18)]. Linear covariance is sufficient for the Newtonian limit and it is obvious without calculation that the $\vartheta$ condition has that much covariance. Consequently, Einstein restricted the allowed coordinate transformations to the $\vartheta$ transformations under which the $\vartheta$ expression transformed as a tensor.

On page 19L, Einstein reduced the Ricci tensor to the weak-field equations (1) by the harmonic coordinate condition (2). Einstein's 1912 static gravitational field in the form of equation (22) does not satisfy the harmonic coordinate condition. So the Ricci tensor would appear to Einstein not to reduce to the Newtonian form. Norton advances the hypothesis that "Einstein would have regarded this failure as a major defect – perhaps in itself sufficient basis for Grossmann's claim" from 1913 (see Section 3.1) that "it turns out that in the special case of the infinitely weak, static gravitational field" the Ricci "tensor does *not* reduce to the" Poisson equation (Einstein and Grossmann, 1913, 35-36). According to Norton, "it is not surprising that Einstein should then have continued to search for another generally covariant gravitation tensor and proceeded to arrive (on page 22R) at $T_{il}^{x}$. The coordinate condition associated with this tensor, [the Hertz condition] equation (2a), is satisfied in Einstein's static field, with the [static spatial flat] metric".

The question (why did Einstein choose the Hertz restriction?) impinges on the matter of whether there is evidence that the incompatibility between the harmonic condition and the spatially flat metric that Einstein thought should describe weak static fields played a role in his search for gravitational field equations. Unlike Norton, Renn and Janssen hold that "there is no evidence that the incompatibility between the harmonic condition and Einstein's prejudice about the form of the metric for weak static fields played a role at this juncture" (Renn and Janssen 2007, 879).

Norton has then advanced the following conjecture: Einstein, however, abandoned the use of coordinate conditions because of the same error committed in the context of the hole argument. Here I explain Norton's conjecture in my own words. We should recall that Norton's conjecture for why Einstein rejected the November tensor $T_{il}^x$ is based on strongly held misconceptions or "prejudices". The November tensor $T_{il}^x$ is covariant under unimodular transformations. Norton's conjecture is that one particular unimodular transformation comes to special prominence in the pages immediately following the proposal of $T_{il}^x$ (pages 42L–43L): Einstein examined the transformation from inertial to uniformly rotating coordinates, equation (15). The simple reading of this transformation is a change of coordinates; starting with the Minkowski metric in standard coordinates, equation (20), one arrives at a metric in rotating coordinates, the rotation metric, equation (15). The metrics (20) and (15) are both solutions of the November tensor $T_{il}^x$ (which is covariant under unimodular transformations).

Einstein, however, specified the coordinate system by imposing the Hertz condition (2a) on the tensor $T_{il}^x$. He thus "created" a hole construction (the Hertz coordinate systems + a hole). The Hertz condition (2a) reduces the tensor $T_{il}^x$ to a reduced November tensor in the Hertz coordinate systems. The two metrics (20) and (15) are related by the transformation from inertial to uniformly rotating coordinates. Einstein would expect that both (20) and (15) would be admissible as solutions in the Hertz coordinate systems (outside the hole). He would expect that both (20) and (15) would be solutions of the reduced November tensor $T_{il}^x$ in the Hertz coordinate systems. But this can only obtain if the Hertz coordinate condition is covariant under the transformation from inertial to uniformly rotating coordinates. The Hertz condition, however, failed to have sufficient covariance, it conflicted with (15).

In a viable coordinate-independent metric theory, the coordinate condition used and the resulting reduced gravitational field equations will still exhibit broad covariance (a hole would not be "created"), including covariance under the transformation from inertial to uniformly rotating coordinates, so that they admit (15) as a solution. Thus, under the normal understanding of coordinate conditions, Einstein would have no reason to check the covariance of the Hertz condition. But, Norton suggests that if Einstein accords independent physical reality to coordinate systems introduced by coordinate conditions, then the natural outcome is to check its covariance. This would have forced Einstein to restrict the covariance of his theory, even had he known full well how to use coordinate conditions in the modern sense. Perhaps this explains why Einstein checked the covariance of the Hertz condition. From page 22R onwards Einstein limited the covariance of his theory to the covariance of the coordinate restrictions and he turned to using coordinate restrictions.

Norton's conjecture is that, this explanation makes clear why Einstein abandoned the use of coordinate conditions so precipitously, why he would have judged the calculation concerning the Newtonian limit of page 22R to be a failure and why he acquiesced so readily to the gravely restricted covariance of the *Entwurf* theory. The *Entwurf* theory does not require coordinate conditions for the recovery of the Newtonian limit. Its gravitation tensor already has the Newtonian form. So merely presuming a weak field of form indirectly introduces enough of a restriction on the coordinate system to allow recovery of the Newtonian limit.

My comments on Norton's conjecture: Norton suggests that Einstein knew about the possibility of coordinate conditions, that he used them in the *Zurich Notebook* and then abandoned them (from page 22R onwards) in favor of the use of coordinate restrictions. Although on pages 19L-20L Einstein did not check the covariance of the harmonic condition (2) and the Hertz condition (2a) in the way that he checked the covariance of the Hertz and $\vartheta$ conditions, it seems most unlikely that Einstein would have rejected the modern notion of coordinate conditions in 1912. It seems probable that Einstein used throughout the *Zurich Notebook* coordinate restrictions because on pages 19L-20L he imposed the harmonic coordinate condition (2) alone, which ensured both the elimination of unwanted second-order derivative terms from the Ricci tensor and the vanishing of the divergence of the stress-energy tensor. Einstein thus employed the same restriction to eliminate unwanted second-order derivative terms from the Ricci tensor and guarantee energy-momentum conservation (see example at the end of Section 3.2).

## 3.5. The Hole Argument

Norton's conjecture for why Einstein rejected the November tensor $T_{il}^x$ is based on "prejudices". Norton has suggested that Einstein abandoned the use of coordinate conditions because of the same error committed in the context of the hole argument, that is to say, there are preferred structures in spacetime that can distinguish one set of coordinates from another. Norton's suggestion leads to the possible conjecture that in 1912-1913, Einstein might have already arrived at a preliminary idea of the hole argument or by that time, he had had the tacit assumptions implicit in the hole argument. Perhaps in the winter of 1912-1913, Einstein rejected the tensor $T_{il}^x$ because he had already possessed an immature version of the hole argument? I will now check this conjecture.

Janssen points out that the earliest version of the hole argument might be found at the bottom of the second page of what appears to be Besso's notes of discussions with Einstein, dated from August 1913 (what scholars call, "Besso's memo") (Janssen 2007, 821). In June 1913, Besso visited Einstein in Zurich, and they both tried to solve the new *Entwurf* field equations to find the perihelion advance of Mercury in the field of the static Sun, in the *Einstein-Besso manuscript*. Before Einstein and Besso finished their joint project of calculations in June 1913, Besso had to leave Zurich and return home to Gorizia where he lived at the time. Besso visited Einstein again in August 1913. During this time, they continued to work on the project, and added more pages to the *Einstein-Besso manuscript*. Besso wrote in his *memo* (dated August 28, 1913) about the failure of realizing general covariance with the gravitational *Entwurf* field equations. He divided space-time into regions with and without matter. Besso imagined a central mass to be surrounded by empty space and wondered whether the solution for the metric tensor was, in this case, determined uniquely for the empty region. What Besso described, could be called an inverted hole argument or a hole argument without a hole. This formulation suggests that the argument concerns the uniqueness of the metric field of the Sun, which Einstein and Besso calculated in their attempt to account for the perihelion anomaly of Mercury on the basis of the *Entwurf* theory. Janssen conjectures that the above version of the hole argument found in the *Besso memo* could be an embryonic version of the hole argument that Einstein had just told Besso about during his August visit in Zurich (Janssen 2007, 821).

If we can provide strong evidence in support of this conjecture, it may indicate with a fairly high probability that, Einstein could have rejected the November tensor $T_{il}^x$ also because he had already possessed an initial version of the hole argument. This could yield important indirect support for Norton's conjecture: According to Norton, Einstein accorded a physical reality to coordinate systems independent of the metric field defined on them. This deep-seated misconception underlies the hole argument and Einstein's rejection of the tensor $T_{il}^x$ in the winter of 1912-1913.

However, the conjecture that Einstein possessed the hole argument during August 1913 seems highly unlikely because in two August 1913 letters to Lorentz Einstein did not yet mention the hole argument. The two letters that he wrote to Lorentz on August 14 and 16, 1913, detail his dissatisfaction with the *Entwurf* field equations. He told Lorentz that, unfortunately, the gravitation equations do not possess the property of general covariance. Einstein felt that the theory refuted its own starting point. He then suggested a solution to his problem: Assuming conservation of momentum and energy, his gravitational field equations were never absolutely covariant. If we restricted the choice of the reference systems, with respect to which the law of momentum and energy conservation holds, then general linear transformations remained the only right choice (Einstein to Lorentz, August 14, 16, 1913, *CPAE 5*, Doc. 467, 470).

According to Renn and Sauer, the version of the hole argument found in the *Besso memo* could have originated with Besso, and might have been Besso's idea after all. Besso could have presented it in his *memo* and brought this idea to the meeting with Einstein only by the end of August 1913. Then, after the meeting in August 1913, the idea could have transformed into Einstein's hole argument. Renn and Sauer point out that Einstein's reaction to Besso's description of an inverted hole argument seems to be preserved in a text written below. He apparently found the inverted hole argument "of no use" (Renn and Sauer 2007, 239-241). In this we see difficulty to Norton's conjecture. If indeed Einstein found the initial version of the hole argument "of no use", then the difficulty now lies in how to reconcile with this evidence Norton's suggestion.

There is another evidence, however, that might speak against Norton's conjecture. Stachel has discovered that before November 1913, Einstein did not possess the hole argument, not even an initial idea of that argument. On September 9, 1913, Einstein gave a lecture in Frauenfeld, titled, "Physical Foundations of a Theory of Gravitation". In this lecture, Einstein could not yet deal with the fact that his gravitational field equations were not covariant with respect to arbitrary, but only covariant with respect to linear transformations. Two weeks later, on September 23, 1913,

Einstein attended the eighty-fifth Congress of the German Natural Scientists and Physicists in Vienna. There he presented another talk, "On the Present State of the Problem of Gravitation" concerning his *Entwurf* theory. A text of this lecture with the discussion was published in the December volume of the *Physikalische Zeitschrift*. Einstein rewrote, in Section §7 of his Vienna paper, the gravitational field equations he had obtained in the *Entwurf* 1913 paper (Einstein 1913, 1258-1259). In regards to the field equations, there was little new in the Vienna talk. Before presenting the field equations he said that, the whole problem of gravitation would be solved satisfactorily, if one were able to find field equations covariant with respect to any arbitrary transformations that are satisfied by the metric tensor components that determine the gravitational field itself. However, he had not succeeded in solving that problem in this manner. Einstein had added a footnote, which indicated that he did arrive at some new idea. He said that in the last few days, he had found proof that such a generally covariant solution could not exist at all (Einstein 1913, 1257). This footnote, however, appeared in the printed version of the Vienna lecture – published on December 15, 1913. Hence the footnote could be added only later, after September 1913. If it was added just before December 1913, a short time earlier, in November 1913, Einstein had an ingenious idea. Indeed, on November 2, 1913, Einstein told his assistant, Ludwig Hopf that he was now very happy with his gravitation theory. The fact that the gravitational equations were not generally covariant, which bothered him up till then, has proved to him now to be unavoidable. Einstein explained to Hopf that it could easily be proved that a theory with generally covariant equations could not exist (Einstein to Hopf, November 2, 1913, *CPAE 5*, 1913, Doc. 480; Stachel, 1989a, 308-309); the proof was the hole argument. Thus, at the present, incomplete stage of our knowledge, it seems that Norton's conjecture, though ingenious, is not substantiated.

## 4. A Combined Conjecture

Different scholars view and interpret the *Zurich Notebook* and Einstein's rejection of the November tensor $T_{il}^x$ from their own angles. I therefore endeavor to reformulate the conjectures offered by scholars in the form of the following combined conjecture.

Renn and Janssen have suggested that Einstein was unable to eliminate all unwanted terms from the November tensor $T_{il}^x$ and to satisfy the conservation of energy-momentum. He employed the same restriction, the Hertz restriction, to reduce the field equations to the Newtonian limit in the case of weak static fields, guarantee energy-momentum conservation, and allow the metric for Minkowski space-time in rotating coordinates. These are three separate problems and they are easily solved separately, but in the *Zurich Notebook* they had become entangled with one another. The entanglement was the result of Einstein's use of coordinate restrictions. He used coordinate restrictions "in a one-size-fits-all fashion": The same coordinate condition suits all problems. Einstein was thus unable to recognize that the November tensor $T_{il}^x$ allows transformations to rotating coordinates.

That which has been suggested by Pais and Norton was taken into consideration by Renn and Janssen. In 1982 Pais suggested that in 1913 Einstein and Grossmann did not yet possess coordinate conditions in the modern sense. Pais, however, did not examine the *Zurich Notebook*. One should recall that Norton, who had already studied the *Zurich Notebook* in 1984, has seen it in quite opposite terms. Firstly, according to Norton, the rotation metric does not satisfy the Hertz restriction. Einstein expected the November tensor $T_{il}^x$ to hold in the case of the rotation metric. On page 22R he imposed the Hertz restriction on the tensor $T_{il}^x$ and recovered the reduced November tensor (i.e. the tensor $T_{il}^x$ in unimodular coordinates satisfying the Hertz restriction) that could lead him to Newton's law of gravitation. He dealt with the rotation metric in a manner that suggests that he expected the November tensor $T_{il}^x$ to retain this form in rotating coordinate systems after the application of the Hertz restriction. The Hertz restriction does not hold in this case, an argument for Einstein against the Ricci tensor. Secondly, Norton suggests that one might say that from page 22R onwards Einstein imposed coordinate restrictions because the field equations based on the reduced tensor $T_{il}^x$ are covariant under those unimodular transformations that preserve the Hertz restriction. Before page 22R, however, he had used coordinate conditions. In an attempt to make sense of Norton's above remarks Renn first hit upon the distinction between coordinate conditions and what he has come to call "coordinate restrictions".

It seems to me, though, that the available evidence suggests that throughout the *Zurich Notebook* Einstein was not using modern coordinate conditions. An example may serve to illustrate this (see Section 3.2): On page 20L Einstein wrote the two conditions, the Hertz condition, and the condition on the trace of the weak-field metric. He then, however, crossed them out because the combination of these two conditions caused problems: The trace of the weak-field metric was constant. In order to avoid this problem, Einstein modified the weak-field equations, and added a term on the right-hand side of these equations with the trace of the stress-energy tensor. The second term on the right-hand side was introduced in such a way that it partially solved the problem. However, he crossed out these weak-field equations and again modified the weak-field equations from pages 19L and 19R, in such a way that he imposed the harmonic coordinate condition alone ("in a one-size-fits-all fashion"), which ensured both the elimination of unwanted second-order derivative terms from the Ricci tensor *and* the vanishing of the divergence of the stress-energy tensor. Performing these two steps seems to support Renn's and Janssen's conjecture: these two steps had become entangled with one another. The entanglement was the result of Einstein's employing the same restriction (the harmonic restriction) to eliminate unwanted second-order derivative terms from the Ricci tensor and guarantee energy-momentum conservation.

There is indeed another explanation of why Einstein abandoned the November tensor $T_{il}^x$ and chose the *Entwurf* field equations. The $\vartheta$ expression conflicted with the Minkowski metric in uniformly rotating coordinates (see Section 2). Following Norton's hypothesis, Janssen and Renn write that in 1912, Einstein gave up the tensor $T_{il}^x$ because the Minkowski rotation metric did not satisfy the Hertz restriction and one may also add the $\vartheta$ restriction.

We may also conjecture that Einstein rejected the November tensor $T_{il}^x$ in 1912-1913 because in the *Zurich Notebook* he conceived the restriction to the $\vartheta$ transformations an essential feature of the gravitational theory and not a feature common to a particular representation of a gravitation theory. In Section 2 I showed that scholars have found that on page 24L, Einstein tried to interpret the components of the contravariant $\vartheta$ metric in terms of inertial forces in rotating frames of reference, just as one would do for the ordinary Minkowski metric in rotating coordinates. Given this, it is clear that Einstein was guided by the equivalence principle. Although the contravariant $\vartheta$ metric could be interpreted in terms of centrifugal forces just as the Minkowski rotation metric, one encounters a major problem with this interpretation, namely that Einstein used the wrong expression for the covariant stress-energy tensor. Scholars have concluded that it is not entirely clear what conclusion Einstein drew from this calculation; but one thing is certain, he finally abandoned the idea of the $\vartheta$ restriction (and its siblings) and subsequently he gave up the November tensor $T_{il}^x$.

In 1915 it might have been just the other way round. Einstein's above earlier experience with the $\vartheta$ expression and its modifications, documented in the *Zurich Notebook*, might have helped him to find his path from the *Entwurf* theory to the first November paper. According to Renn and Sauer, in the *Zurich Notebook*, Einstein had already explored the relation between tensors, the Ricci tensor, $T_{il}$ expressed in terms of the Christoffel symbols and those expressed in terms of the derivatives of the metric, in the context of studying the $\vartheta$ coordinate restriction.

Clearly heuristic principles are likely to be a most important factor in Einstein's rejection of the November tensor $T_{il}^x$. Renn and Sauer have also suggested that Einstein had to solve the incompatibility problem between the correspondence and conservation principles. He changed the left hand side of the field equations and solved the conflict between the two above principles, but had to check whether the conservation principle was fully satisfied. Once this issue was settled the correspondence principle caused problems with the theory of static gravitational fields (with Einstein's conception of the metric of static gravitational fields which has to be spatially flat). At the end the match between the correspondence principle and the conservation principle was achieved at the expense of the generalized principle of relativity. At some stage, thus, Einstein appeared to somewhat forgot a little from the generalized principle of relativity; the starting point of his research project. Meanwhile he had developed many tools that would finally lead him to the goal of generally covariant field equations. However, at this stage he was not quite sure about the most important principles his novel theory should fulfill. He found it difficult to establish a match between the equations and the principles.

According to Renn and Sauer, Einstein constantly oscillated between two strategies, the physical strategy (how to cause the Newtonian limit to appear and guarantee that the field equations be compatible with energy-momentum conservation) and the mathematical strategy (constructing a generally covariant candidate for the left-hand side of the field equations).

In the *Zurich Notebook* Einstein started from mathematics (i.e. mathematical strategy). He first tackled relativity and equivalence. He then arrived at general covariance on page 22R, and then moved on to correspondence (Newtonian limit) and conservation of momentum and energy. Afterwards it was just the other way round (from pages 24R onwards). He gave up mathematics (i.e. mathematical strategy) and established field equations – that would lead him to the so-called *Entwurf* field equations – while starting from physics (i.e. physical strategy). He first tackled correspondence and conservation and then relativity and equivalence, and lost general covariance. The interplay of the four heuristic principles with the new tools of absolute differential calculus of 1912 that Einstein was exploring governed the form of the field equations he was finally left with at the end of the *Zurich Notebook*. Consequently, at the end of the day Einstein's conditions overdetermined his research between 1912 and 1913.

Between 1913 and 1914 Einstein adopted a physical strategy and developed the *Entwurf* theory. He practiced his mathematical skills in the *Zurich Notebook* and elaborated the 1913 *Entwurf* theory along the mathematical strategy. In 1914 he thus derived the *Entwurf* field equations along the mathematical strategy. In October 1915 Einstein realized that he had derived the non-covariant *Entwurf* field equations along the mathematical strategy, which is naturally suitable for generally-covariant field equations. After abandoning the *Entwurf* theory Einstein returned to the November tensor $T_{il}^x$ he had derived along the mathematical strategy three years earlier in the *Zurich Notebook*. Renn and Janssen have suggested that the *Entwurf* theory provided Einstein with the scaffolding for solving the problems with the Newtonian limit and conservation of momentum and energy. In the November 4, 1915 paper coordinate restrictions were turned into coordinate conditions; then the relation between covariance and energy–momentum conservation was inverted from conservation restricting covariance to covariance guaranteeing conservation. Janssen and Renn argue that this was thus a success of the physical strategy.

Norton suggests that towards the end of 1915 Einstein chose the mathematical strategy and, writing down his field equations, found himself rapidly propelled towards a theory that satisfied all the requirements and fulfilled his "wildest dreams". Through November 1915, Einstein was in intense competition with the mathematician David Hilbert. Norton demonstrates that Einstein consciously chose to allow the natural mathematical constructions of tensor calculus to guide him and rapidly bring him to the final result, and so the final field equations were communicated by an exhausted Einstein at the end of the month. Norton suggests that at the last minute Einstein reversed his approach and adopted the mathematical strategy. Moreover, says Norton, even at this last moment, Einstein was quite disparaging and even scornful of the heuristic power of mathematical simplicity. Norton argues that Einstein was indifferent and at times even derisive of considerations of mathematical simplicity early in his career. His disparaging attitude towards mathematics was only weakened temporarily in the winter of 1912-1913 by the need to proceed within a context more mathematically sophisticated than any in which he had worked before.

However, Einstein's indifference to mathematical simplicity persisted up to and through the years in which he worked on completing the general theory of relativity.

I would like to add that the competition between Einstein and Hilbert culminated in November of 1915 when Einstein adopted a mathematical approach and Hilbert adopted a physical approach: in the November 20 proofs of Hilbert's paper (which did not contain a generally covariant theory) Hilbert based his assertion on a slightly more sophisticated version of Einstein's hole argument against general covariance (after Einstein had silently dropped it), which he would eventually drop later when he would publish his paper in March 1916. Thus, scholars have already shown that when working on his November papers, Einstein was influenced from the current discussions with Hilbert; and when working on Einstein's theory, Hilbert was influenced from the current discussions with Einstein.[4]

It seems to me that in October-November 1915 both strategies, the physical and the mathematical, came into play. Between 1912 and 1915, Einstein moved back and forth between the two strategies and among several heuristic principles (correspondence, conservation of momentum-energy, generalized principle of relativity, and equivalence) in the course of the development of his field equations. That Einstein attached immense importance to three problems, the Newtonian limit, energy-momentum conservation, and rotation, but to use Norton's adage, "in solving one problem, however, we have created another", is amply demonstrated in many pages of the *Zurich Notebook*.

Renn and Janssen have suggested that in November 1915, Einstein was able to disentangle the problem of energy-momentum conservation from the problem of recovering the Newtonian limit, and he also decoupled the latter from the problem of rotation. He came back to field equations based on the November tensor $T_{il}^x$ when he understood that they can be made both compatible with energy-momentum conservation and the Newtonian limit: He imposed a coordinate restriction only for energy-momentum conservation and the Hertz coordinate condition was added to the field equations (specifying the coordinate system) to obtain the Poisson equation for weak static fields. Einstein used the exact same mathematical formula (2c), but now interpreted as a coordinate condition (2b) rather than a coordinate restriction (2a). Right after using the Hertz condition to demonstrate that his field equations based on the November tensor $T_{il}^x$ have the correct Newtonian limit to first approximation, Einstein briefly demonstrated that those allow transformations to rotating coordinates. Einstein's letter to David Hilbert from November 18, 1915 (quoted in the introduction) seems to support this conjecture.

Norton has argued that Einstein had already used the exact same mathematical formula (2b) on page 19R of the *Zurich Notebook* and subsequently on page 22R he switched to (2a). Norton provides concrete evidence from page 19L. According to Norton, when the harmonic condition was introduced on page 19L, it was used as a coordinate condition, with Einstein perhaps reserving the possibility of using it as a coordinate restriction. Einstein did not check the covariance of the harmonic condition on pages 19L–21R in the way that he checked the covariance of the Hertz restriction on page 22R. From page 22R onwards Einstein limited the covariance of his theory to the covariance of the coordinate restrictions because he accorded independent physical reality to coordinate systems (even had he known full well how to use coordinate conditions). Renn, Janssen, and Sauer, broadly hold that the terms in the Ricci tensor with unwanted second-order derivatives were eliminated by imposing the harmonic coordinate restriction. Einstein extracted from the Ricci tensor weak-field equations that satisfy the correspondence principle. He then checked whether these field equations and the harmonic coordinate restriction are compatible with his other heuristic requirements (i.e. conservation of energy momentum).

---

[4] See Renn and Stachel 2007.

I suppose that the evidence (presented further above) supports the prevailing view of scholarship presented here. It does not mean that the view adopted by the majority of scholars will finally win out.

Surely there were also several prejudices (using Norton's phraseology) or *Ansatz*s (using Stachel's memorable adage) that were leading Einstein away from the November tensor $T_{il}^x$; these misconceptions prevented him from achieving a comprehensive view of this tensor. Norton and Stachel have contributed to a clearer understanding of Einstein's misconceptions in the *Zurich Notebook*. They argue that Einstein's rejection of the Ricci tensor may have also resulted from Einstein's long-held belief that the static gravitational fields are inconsistent with the Ricci tensor. Einstein thought that the same type of metric tensor degeneration appears in special relativity and in static gravitational fields, only that in the latter case $c^2$ is a function of the spatial coordinates. In the weak-field approximation, the spatial metric of a static gravitational field must be flat.

According to Norton, Einstein considered the flat spatial metric of a static gravitational field in the *Zurich Notebook* on page 20L and after his examination of the Ricci tensor in harmonic coordinates (this example is mentioned in Section 3.1 and further examined in Section 3.2). Such a flat spatial metric does not satisfy the harmonic coordinate condition. So the Ricci tensor would appear to Einstein not to reduce to the required Newtonian limit in the case of weak, static field. Einstein would have regarded this failure as a major defect and he thus continued to search for another generally covariant tensor and proceed to arrive at the November tensor $T_{il}^x$. The Hertz condition (2a), says Norton, is satisfied in Einstein's static field, with the flat spatial metric. Renn and Janssen do not agree with Norton. They argue that there is no evidence that the above incompatibility between the harmonic condition and Einstein's prejudice about the form of the metric for weak, static fields played a role in Einstein's rejection of the Ricci tensor. Moreover, they doubt whether Einstein was even aware of this incompatibility at the time of the *Zurich Notebook*.

Norton then goes on to explain that Einstein's explorations were based on the principle of equivalence, which asserted that a transformation to uniform acceleration in a Minkowski spacetime yielded a homogenous gravitational field. If one transforms from the metric of special relativity to a coordinate system in uniform acceleration, the metric reverts to a form Einstein associated with a homogeneous gravitational field, which has the form of the flat spatial metric of a static gravitational field. Thus, Einstein's mistake, in 1912 and 1913, was that he considered the spatial flatness of the above metric of a homogeneous gravitational field a property of all static fields.

According to Stachel, Einstein expected that the Ricci tensor should reduce in the limit of weak-fields to his static gravitational field theory from 1912 and then to the Newtonian limit, if the static spatial metric is flat. Stachel says that this statement appears to have led Einstein to reject the Ricci tensor, and fall into the trap of the *Entwurf* limited generally covariant field equations. Stachel also mentions Einstein's and Besso's calculations of the perihelion of Mercury. In 1913 Einstein calculated the metric field of the Sun using the *Entwurf* vacuum field equations in a first-order approximation and found that the static spatial metric was flat. Two years later, he calculated the metric field of the Sun using the November 1915 vacuum field equations in a first-order approximation and discovered that the static spatial metric need not be flat.

Einstein's flat spatial metric of a static gravitational field was incompatible with field equations containing a term with the trace of the stress-energy tensor of matter, equations (41), (41a) and (44). Norton has suggested that Einstein had added such a trace term to the field equations of November 1915, equation (4c), but only after he had discovered – in the course of his calculations of the advance of the motion of the perihelion of Mercury in November 18 – that the static spatial metric need not be flat. Thus, we can speculate that Einstein could not return to equations (41), (41a), (43) and (44) and find the exact equations corresponding to these weak-field equations before November 18, 1915: as long as he was stubbornly stuck to his *Ansatz*, he could only return to the November tensor $T_{il}^x$. According to Stachel and Norton, shortly *after* Einstein had returned to the November tensor $T_{il}^x$, he renounced his prejudice or *Ansatz* of flat spatial metric of a static gravitational field. In other words, in November 4, 1915 Einstein returned to the November tensor $T_{il}^x$, but he had not yet renounced the above *Ansatz*. He thus did not yet solve the problem of obtaining the Newtonian limit from his new theory.

Norton, however, suggests the possibility of another misconception: Einstein accorded a physical reality to coordinate systems. According to Norton, this deep-seated misconception stood in the way of returning to the November tensor $T_{il}^x$: It underlies the hole argument and also Einstein's rejection of the tensor $T_{il}^x$ in the winter of 1912-1913.

In Section 3.5 I have mentioned the *Besso memo*: Janssen points out that the earliest version of the hole argument might be found in the *Besso memo* dated from August 1913. He conjectures that this version of the hole argument found in the *Besso memo* could be an embryonic version of the hole argument that Einstein had just told Besso about during his August visit in Zurich. However, Renn and Sauer point out that the version of the hole argument found in the *Besso memo* could have originated with Besso, and might have been Besso's idea after all because Einstein's reaction to Besso's description of the so-called hole argument seems to be preserved in a text written below. He apparently found this hole argument "of no use".

If Einstein indeed found this argument of no use, then it seems likely that it could not be the reason why he rejected the November tensor $T_{il}^x$. The difficulty lies in how to reconcile with this evidence Norton's conjecture.

The conjecture put forward by Norton that Einstein abandoned the use of coordinate conditions because of the same error committed in the context of the hole argument – has not found a good deal of support among other scholars. "The evidence available makes it", in Renn's view, "implausible that this was indeed Einstein's pitfall in early 1913". In 2005 Renn was rather doubtful: "More generally speaking, it seems that searching for the fatal error supposedly responsible for the pitfall of early 1913 may be something like the hunt for the white elephant in Mark Twain's famous story" (Renn 2005, 45-46) Or like *The Hunting of the Snark* in Lewis Carroll's great nonsense poem.

In 1913 Einstein had built for himself a spacious castle. Very soon he seemed to feel much pleasanter at his new home, the *Entwurf* theory. Take a minute to think about it. According to Stachel, in 1913 Einstein formulated the *Entwurf* field equations based on the criteria that the field equations should generalize Poisson's equation, be invariant at least under linear transformations and the conservation laws for energy and momentum should follow from the gravitational field equations. He also demonstrated how Newton's law of gravitation follows from the linear approximation of the *Entwurf* field equations for the static case. He explicitly noted that the spatial metric remains flat in this approximation. He developed the hole argument against general covariance of the *Entwurf* field equations. Finally, Einstein checked the *Entwurf* field equations using the rotation metric. He thought that his expression vanished for the rotation metric, a necessary condition for the rotation metric to be a solution of the *Entwurf* vacuum field equations. This was a mistake that Einstein would only discover in October 1915. The *Entwurf* theory, says Norton, does not require coordinate conditions for the recovery of the Newtonian limit. Its gravitation tensor already has the Newtonian form. So merely presuming a weak field of form indirectly introduces enough of a restriction on the coordinate system to allow recovery of the Newtonian limit. Stachel adds that deriving Newton's law from the linear approximation of the *Entwurf* field equations for the static case (for which the spatial metric is flat), demonstrating that the rotation metric was a solution of his *Entwurf* field equations and the hole argument prevented Einstein from changing his beliefs about the Newtonian limit and accepting the November tensor until November 1915.

Renn and Janssen have suggested that it took Einstein three years to arrive at the realization that the November tensor $T_{il}^x$ is not incompatible with Newton's law of gravitation because the *Entwurf* theory had a crucial role as scaffolding for building an arch or a bridge between physics and mathematics. On its basis, Einstein first calculated the Mercury perihelion motion, and then worked out the preliminary mathematical structure of general relativity. The Mercury calculation eventually helped him to solve the problem of the Newtonian limit. Einstein and Grossman set up a mathematical variational formalism for the *Entwurf* field equations, from which he could eventually derive the conservation of energy and momentum. By November 1915, Einstein had the scaffold torn down only to be led back to his good old 1912 November tensor $T_{il}^x$.

## 5. Conclusion

In the introduction I posed three questions:

1) Why did Einstein reject the November tensor $T_{il}^x$ in 1912-1913, only to come back to it in November 1915?

2) What did Einstein mean when he said that it was hard for him to recognize that the November tensor is a simple and natural generalization of Newton's law of gravitation?

3) Why did it take him three years to arrive at the realization that the November tensor is not incompatible with Newton's law? I presented several conjectures raised thus far by several historians of physics to answer these questions.

My formulation (in sections 3.1-3.4) of the conjectures proposed by different scholars is an attempt to provide a summary from my own perspective. I believe that the combined conjecture provides a novel perspective to the problem because I attempt to combine several of the proposed conjectures into one coherent explanation:

Firstly, Stachel's and Norton's conjectures that Einstein's rejection of the Ricci tensor may have resulted from Einstein's long-held belief that the static gravitational fields are inconsistent with the Ricci tensor.

Secondly, Renn's and Janssen's conjecture that in the *Zurich Notebook* and in the *Entwurf* theory, one and the same coordinate restriction reduces the field equations to the Newtonian limit, ensures energy-momentum conservation, and demonstrates that the metric for Minkowski spacetime in rotating coordinates is a solution of the field equations. The three problems can be disentangled by switching from coordinate restrictions to coordinate conditions. Einstein seems to have made this switch when he adopted the square root of the negative determinant of the metric tensor equal to 1 as a postulate (restriction to unimodular transformations). Energy-momentum conservation was then fulfilled; thereafter Einstein proceeded by applying the Hertz coordinate condition and was able to obtain the Poisson equation, after which he solved the problem of rotation.

Thirdly, Renn's and Sauer's conjecture that Einstein had to solve the incompatibility problem between the correspondence and the conservation principles. In the *Zurich Notebook* Einstein first tackled two heuristic principles, relativity and equivalence. He then arrived at general covariance on page 22R and moved on to the correspondence principle and conservation of momentum and energy principle. Afterwards it was just the other way round (from pages 24R onwards). He first tackled correspondence and conservation and then relativity and equivalence, and lost general covariance. The interplay of the four heuristic principles with the new tools of absolute differential calculus that Einstein was exploring governed the form of the field equations he was finally left with at the end of the *Zurich Notebook*.

Fourthly, the group of scholars who have been studying Einstein's *Zurich Notebook* and papers over the past few years have concluded that in 1912-1913 Einstein constantly oscillated between two strategies, the physical strategy and the mathematical strategy. As discussed above in Section 4, there is some debate among scholars whether Einstein had ever fully given up the mathematical strategy, or whether he had ever fully adopted it. Generally speaking, it seems that in 1913 Einstein gave up the mathematical strategy and rejected the November tensor. He did not extract a candidate for the left-hand side of the field equations from the Ricci tensor but established the *Entwurf* field equations, while starting from the physical strategy, the requirement of the conservation of momentum and energy. Finally, in November 1915 he returned to the mathematical strategy and to the November tensor $T_{il}^{x}$.

According to the theory put forward by scholars, Einstein's difficulty with the November tensor $T_{il}^{x}$ had lain in misconceptions, tacit assumptions, entanglement among three problems, interplay of physics (heuristic principles) with mathematics, and imposing coordinate restrictions.

Finally, I believe that the key to the problem (Why Einstein rejected the November tensor $T_{il}^{x}$ in 1912 and came back to it in 1915) also lay in the conflict between the $\vartheta$ expression and the Minkowski metric in uniformly rotating coordinates. Einstein's experience with the $\vartheta$ expression, documented in the *Zurich Notebook*, seems to have led him to reject the tensor $T_{il}^{x}$ and it might have also helped him to find his path from the *Entwurf* theory back to the November tensor. This conjecture combines some ideas of Renn, Janssen, and Sauer.

Stachel points out that he believes that the most one can hope to do in discussing Einstein's route to relativity is to construct a plausible conjecture. Such a conjecture will be based upon a certain weighting of the scanty evidence we possess, based upon certain methodological hypotheses, as well as the imagination of the conjecturer. There are bound to be differences of opinion in these matters. All one can demand is that it be made clear on what methodological hypotheses a conjecture is based, and a demonstration that the conjecture is in accord with the available evidence when the latter is weighted in accord with these hypotheses (Stachel, 1989b, 158).

Indeed, several historians have formulated plausible conjectures to explain Einstein's rejection of the Ricci tensor. They provided important but scant evidence in support of their conjecture. Obviously, historians often disagree with one another. However, at the end of the day, from their point of view, they all provided fragmentary pieces of evidence in support of their conjecture.

Finally, I endeavored to reformulate the conjectures raised thus far by several scholars in a form of a combined conjecture. I argue that the combined conjecture is plausible and maybe the best answer we currently have to the three questions.

*Acknowledgments*: I wish to thank Prof. John Stachel from the Center for Einstein Studies in Boston University for sitting with me for so many days discussing Einstein's general relativity and its history.